\documentclass[english,onecolumn]{IEEEtran}
\usepackage[T1]{fontenc}
\usepackage[latin9]{inputenc}
\usepackage{array}
\usepackage{multirow}
\usepackage{amsmath}
\usepackage{amsthm}
\usepackage{amssymb}
\usepackage{graphicx}
\usepackage{esint}

\makeatletter

\providecommand{\tabularnewline}{\\}

\theoremstyle{plain}
\newtheorem{thm}{\protect\theoremname}
\theoremstyle{definition}
\newtheorem{defn}[thm]{\protect\definitionname}
\theoremstyle{plain}
\newtheorem{prop}[thm]{\protect\propositionname}
\theoremstyle{definition}
\newtheorem{example}[thm]{\protect\examplename}

\usepackage{mathrsfs}

\usepackage{colortbl}
\definecolor{lightgray}{rgb}{0.9,0.9,0.9}
\definecolor{lightred}{rgb}{1,0.8,0.8}
\definecolor{lightgreen}{rgb}{0.6,1,0.6}
\definecolor{lightyellow}{rgb}{1,1,0.5}
\definecolor{lightgrey}{rgb}{0.8,0.8,0.8}

\allowdisplaybreaks[1]

\makeatother

\usepackage{babel}
\providecommand{\definitionname}{Definition}
\providecommand{\examplename}{Example}
\providecommand{\propositionname}{Proposition}
\providecommand{\theoremname}{Theorem}

\begin{document}
\title{A Poisson Decomposition for Information and the Information-Event
Diagram}
\author{Cheuk Ting Li\\
Department of Information Engineering, The Chinese University of Hong
Kong\\
Email: ctli@ie.cuhk.edu.hk}
\maketitle
\begin{abstract}
Information diagram and the I-measure are useful mnemonics where random
variables are treated as sets, and entropy and mutual information
are treated as a signed measure. Although the I-measure has been successful
in machine proofs of entropy inequalities, the theoretical underpinning
of the ``random variables as sets'' analogy has been unclear until
the recent works on mappings from random variables to sets by Ellerman
(recovering order-$2$ Tsallis entropy over general probability space),
and Down and Mediano (recovering Shannon entropy over discrete probability
space). We generalize these constructions by designing a mapping which
recovers the Shannon entropy (and the information density) over general
probability space. Moreover, it has an intuitive interpretation based
on the arrival time in a Poisson process, allowing us to understand
the union, intersection and difference between (sets corresponding
to) random variables and events. Cross entropy, KL divergence, and
conditional entropy given an event, can be obtained as set intersections.
We propose a generalization of the information diagram that also includes
events, and demonstrate its usage by a diagrammatic proof of Fano's
inequality.
\end{abstract}

\medskip{}

\section{Introduction}

The analogy between information (or random variables) and sets was
first studied by Hu \cite{ting1962amount}, and utilized in information
diagrams \cite{cover2006elements} and the I-measure by Yeung \cite{yeung1991new,yeung2012first}.
In this analogy, random variables are treated as sets. The entropy
of a random variable corresponds to the (signed) measure of the set.
The joint entropy, mutual information and conditional entropy between
two random variables correspond to the measure of the union, intersection
and difference between the two sets respectively \cite{ting1962amount}.
Information diagrams have been useful mnemonic aids for equalities
and inequalities between entropy and mutual information terms, and
the I-measure \cite{yeung2012first} has been successful as a symbolic
tool for machine proofs of entropy inequalities \cite{yeung1996itip,yeung1997framework,ho2020proving,li2021automatedisit}.

The question on whether the ``information as set'' analogy has a
deeper theoretical foundation, which explains why mutual information
and conditional entropy appear as the measures of the intersection
and difference respectively, was raised by Campbell \cite{campbell1965entropy}.
 While the joint random variable $(X,Y)$ can indeed be considered
as the ``union'' of the information in $X$ and $Y$, with entropy
$H((X,Y))$ coinciding with the joint entropy $H(X,Y)$, the same
cannot be said for intersection and difference. Several notions of
common information has been proposed (e.g. \cite{gacs1973common,wyner1975common,kumar2014exact}),
which can be treated as the random variable that is the ``intersection''
between two random variables $X,Y$, though none of them generally
has entropy $I(X;Y)$. There are also attempts at finding the random
variable that corresponds to the ``difference'' between two random
variables $X,Y$ (e.g. \cite{sfrl_trans,li2017extended}), though
none of them generally has entropy $H(Y|X)$. There are no clear meanings
of the intersection or difference between two random variables, if
the results must also be random variables. The I-measure can only
tell what the measures of the intersection and difference are, but
not what those sets actually correspond to.

To allow all Boolean set operations (union/intersection/difference),
we have to embed the space of random variables onto a larger space
of sets that not only contains random variables. Here we review two
attempts to construct such embedding by Ellerman \cite{ellerman2017logical}
and Down and Mediano \cite{down2023logarithmic}.

Ellerman \cite{ellerman2017logical,ellerman2018logical} studied the
mapping which maps a random variable $X:\Omega\to\mathcal{X}$ to
the set of pairs $(\omega_{1},\omega_{2})\in\Omega^{2}$ that $X$
distinguishes, i.e., $\mathrm{dit}(X)=\{(\omega_{1},\omega_{2})\in\Omega^{2}:\,X(\omega_{1})\neq X(\omega_{2})\}$.
The joint random variable $(X,Y)$ indeed corresponds to the union
of the two sets, i.e., $\mathrm{dit}((X,Y))=\mathrm{dit}(X)\cup\mathrm{dit}(Y)$.
The order-$2$ Tsallis entropy \cite{tsallis1988possible}, which
is the probability $(\omega_{1},\omega_{2})\in\mathrm{dit}(X)$ for
$\omega_{1},\omega_{2}$ i.i.d., is then a measure over such sets
of pairs.\footnote{We remark that \cite{ellerman2017logical} focuses on finite uniform
probability spaces, though it is straightforward to extend the construction
to general probability spaces.}

Recently, Down and Mediano \cite{down2023logarithmic} introduced
the logarithmic decomposition for information, which is based on a
mapping which maps a random variable $X:\Omega\to\mathcal{X}$ to
the set of subsets $B\in2^{\Omega}$ that $X$ can distinguish at
least one pair of elements, i.e., $\mathcal{C}(X)=\{B\in2^{\Omega}:\,|X(B)|>1\}$
(where $X(B)$ is the image of $B$). Then a signed measure over $2^{\Omega}$
is defined via M{\"o}bius inversion such that the Shannon entropy
$H(X)$ coincide with the measure of $\mathcal{C}(X)$. Unlike \cite{ellerman2017logical},
the set $B$ here can have size larger than $2$, which allows the
Shannon entropy (instead of the order-$2$ Tsallis entropy) to be
defined as a signed measure. A limitation is that the M{\"o}bius
inversion construction only works when the sample space $\Omega$
is finite, over which only finitely many non-equivalent random variables
can be defined. Hence, \cite{down2023logarithmic} is subject to the
same limitation as the I-measure \cite{yeung2012first}, which can
only describe finitely many random variables simultaneously. 

In this paper, we provide a mapping $\tilde{\mathrm{G}}$ from a random
variable $X$ to a set $\tilde{\mathrm{G}}(X)$ in a ``generalized
information space'', equipped with a signed measure, called the \emph{Poisson
information measure}, which has an intuitive meaning based on the
arrival time of a Poisson process. The construction can be applied
on an arbitrary (discrete/continuous) probability space, and hence
can be considered as a generalization of \cite{down2023logarithmic}
to general probability spaces (not only finite ones). Notably, we
can also define a mapping $\mathrm{G}$ from an event $E$ to a set
$\mathrm{G}(E)$. We retain the following conventional insights provided
by information diagrams and I-measure (that are also realized by the
mapping in \cite{down2023logarithmic}):
\begin{itemize}
\item A random variable $X$ is mapped to a set $\tilde{\mathrm{G}}(X)$,
and the signed measure of $\tilde{\mathrm{G}}(X)$ is the Shannon
entropy $H(X)$.
\item If $Y$ is a function of $X$, then $\tilde{\mathrm{G}}(Y)\subseteq\tilde{\mathrm{G}}(X)$.
\item The joint random variable $(X,Y)$ is mapped to the set union $\tilde{\mathrm{G}}((X,Y))=\tilde{\mathrm{G}}(X)\cup\tilde{\mathrm{G}}(Y)$.
\item The signed measure of the set difference $\tilde{\mathrm{G}}(Y)\backslash\tilde{\mathrm{G}}(X)$
is the conditional entropy $H(Y|X)$.
\item The signed measure of the intersection $\tilde{\mathrm{G}}(X)\cap\tilde{\mathrm{G}}(Y)$
is the mutual information $I(X;Y)$.
\end{itemize}
In addition, this paper also provides the following new insights:
\begin{itemize}
\item The operation of conditioning a random variable $X$ on an event $E$
correspond to set intersection $\tilde{\mathrm{G}}(X)\cap\mathrm{G}(E)$,
and the signed measure of $\tilde{\mathrm{G}}(X)\cap\mathrm{G}(E)$
is $\mathbb{P}(E)H(X|E)$. Similarly, the signed measure of $\tilde{\mathrm{G}}(X)\cap\tilde{\mathrm{G}}(Y)\cap\mathrm{G}(E)$
is $\mathbb{P}(E)I(X;Y|E)$. Moreover, conditioning on an event corresponds
to thinning the Poisson process by discarding points outside of $E$.
\item If the random variable $X$ induces the partition $E_{1},\ldots,E_{n}$,
then $\tilde{\mathrm{G}}(X),\mathrm{G}(E_{1}),\ldots,\mathrm{G}(E_{n})$
partition the range of $\mathrm{G},\tilde{\mathrm{G}}$. Intuitively,
$X$ captures all information that are not localized in any one of
$E_{1},\ldots,E_{n}$. This provides an intuitive picture of what
``a random variable $Y$ conditioned on a random variable $X$''
means in terms of the information it contains. This also gives a simple
diagrammatic interpretation of the equality $H(Y|X)=\sum_{x}\mathbb{P}(X=x)H(Y|X=x)$. 
\item Moreover, cross entropy and the Kullback-Leibler divergence can also
be given as the set intersection. This gives a simple diagrammatic
interpretation of the equality $I(X;Y)=\sum_{x}\mathbb{P}(X=x)D_{\mathrm{KL}}(P_{Y|X=x}\Vert P_{Y})$. 
\item The Poisson information measure can also recover pointwise measures
of information such as the self-information and the information density.
\end{itemize}
See Table \ref{tab:correspondence} for a list of correspondence between
random variables, events, and set operations. Based on the above insights,
we propose a generalization of the information diagram, called the
\emph{information-event diagram}, which can include both random variables
and events. To demonstrate the usage of this diagram, we give a diagrammatic
proof of Fano's inequality \cite{fano1961transmission}. We introduce
a symbolic tool called the \emph{IE-measure}, which extends the I-measure
\cite{yeung1991new,yeung2012first} to also consider events.

\begin{table}
\begin{centering}
{\renewcommand*{\arraystretch}{1.7}%
\begin{tabular}{|c|c|}
\hline 
\textbf{Concept} & \textbf{Generalized information}\tabularnewline
\hline 
\hline 
Event $E$ & $\mathrm{G}(E)=\bigcup_{k=1}^{\infty}E^{k}$\tabularnewline
\hline 
Empty event $\emptyset$ & Empty $\mathrm{G}(\emptyset)=\emptyset$\tabularnewline
\hline 
Subset $E_{1}\subseteq E_{2}$ & Subset $\mathrm{G}(E_{1})\subseteq\mathrm{G}(E_{2})$\tabularnewline
\hline 
Disjoint $E_{1}\cap E_{2}=\emptyset$ & Disjoint $\mathrm{G}(E_{1})\cap\mathrm{G}(E_{2})=\emptyset$\tabularnewline
\hline 
Intersection $E=E_{1}\cap E_{2}$ & Intersection $\mathrm{G}(E)=\mathrm{G}(E_{1})\cap\mathrm{G}(E_{2})$\tabularnewline
\hline 
\hline 
Random variable $X$ & $\tilde{\mathrm{G}}(X)=\mathrm{G}(\Omega)\,\backslash\,\bigcup_{x\in\mathcal{X}}\mathrm{G}(X^{-1}(x))$\tabularnewline
\hline 
\multirow{2}{*}{Partition $E_{1},\ldots,E_{n}$ of $\Omega$} & Partition $\mathrm{G}(E_{1}),\ldots,\mathrm{G}(E_{n}),\tilde{\mathrm{G}}(X)$
of $\mathrm{G}(\Omega)$,\tabularnewline
 & where $X$ is the RV with induced partition $\{E_{i}\}_{i}$\tabularnewline
\hline 
Constant RV $X=c$ & Empty $\tilde{\mathrm{G}}(X)=\emptyset$\tabularnewline
\hline 
Functional dependency $H(Y|X)=0$ & Subset $\tilde{\mathrm{G}}(Y)\subseteq\tilde{\mathrm{G}}(X)$\tabularnewline
\hline 
Joint RV $Z=(X,Y)$ & Union $\tilde{\mathrm{G}}(Z)=\tilde{\mathrm{G}}(X)\cup\tilde{\mathrm{G}}(Y)$\tabularnewline
\hline 
Conditioning on event $E$ & Thinning the Poisson process by discarding points outside $E$\tabularnewline
\hline 
RV $X$ conditional on event $E$ & Intersection $\tilde{\mathrm{G}}(X)\cap\mathrm{G}(E)$\tabularnewline
\hline 
RV $Y$ conditional on RV $X$ & Difference $\tilde{\mathrm{G}}(Y)\backslash\tilde{\mathrm{G}}(X)$\tabularnewline
\hline 
\hline 
Entropy $H(X)$ & Poisson info. measure $\mathcal{H}(\tilde{\mathrm{G}}(X))$\tabularnewline
\hline 
Self-information $\iota_{X}(X(u))$ & Poisson info. measure $\mathcal{H}_{u}(\tilde{\mathrm{G}}(X))$\tabularnewline
\hline 
Conditional entropy $H(Y|E)$$\,^{*}$ & Measure of intersection $\mathcal{H}(\tilde{\mathrm{G}}(X)\cap\mathrm{G}(E))$\tabularnewline
\hline 
Conditional entropy $H(Y|X)$ & Measure of difference $\mathcal{H}(\tilde{\mathrm{G}}(Y)\backslash\tilde{\mathrm{G}}(X))$\tabularnewline
\hline 
Mutual information $I(X;Y)$ & Measure of intersection $\mathcal{H}(\tilde{\mathrm{G}}(X)\cap\tilde{\mathrm{G}}(Y))$\tabularnewline
\hline 
Information density $\iota_{X;Y}(X(u);Y(u))$ & Measure of intersection $\mathcal{H}_{u}(\tilde{\mathrm{G}}(X)\cap\tilde{\mathrm{G}}(Y))$\tabularnewline
\hline 
Cross entropy $H(P_{X|E},P_{X|F})$$\,^{*}$ & Measure of intersection $\mathcal{H}(\tilde{\mathrm{G}}(X)\cap\mathrm{G}(E,F))$\tabularnewline
\hline 
KL divergence $D_{\mathrm{KL}}(P_{X|E}\Vert P_{X|F})$$\,^{*}$ & Measure of intersection $\mathcal{H}(\tilde{\mathrm{G}}(X)\cap(\mathrm{G}(E,F)\backslash\mathrm{G}(E)))$\tabularnewline
\hline 
\end{tabular}}
\par\end{centering}
\medskip{}

\caption{\label{tab:correspondence}Correspondence between concepts in probability
and information theory, and the corresponding concepts for generalized
information. For the lines marked with asterisks, there is a scaling
factor $\mathbb{P}(E)$ omitted (see Proposition \ref{prop:cond_event}),
and the identity for KL divergence works only when $E\subseteq F$
(see Proposition \ref{prop:kl}).}
\end{table}

\medskip{}

\subsection*{Other Related Works}

For a fixed number of random variables, finding the set of possible
I-measures over the atoms is equivalent to finding the entropic region
\cite{zhang1997non,yeung1997framework,zhang1998characterization}.
While the entries in the entropic region must satisfy the polymatroidal
inequalities \cite{fujishige1978polymatroidal}, this is not a complete
characterization, and inequalities that do not follow from the polymatroidal
inequalities have been given in \cite{zhang1998characterization,makarychev2002new,dougherty2006six,matus2007infinitely}.
The characterization of the entropic region for $4$ random variables
is still open, and various problems about the entropic region have
been shown to be undecidable \cite{li2022undecidabilityaffine,li2023undecidability,kuhne2022entropic}.
Since the IE-measure in this paper is an extension of the I-measure,
problems on the IE-measure would be at least as hard as the corresponding
problems on the I-measure.

In order to study the synergy between a collection of random variables
$X_{1},\ldots,X_{n}$ and another random variable $Y$, Williams and
Beer introduced the partial information decomposition \cite{williams2010nonnegative},
where the mutual information $I(Y;X_{1},\ldots,X_{n})$ is decomposed
into cells with measures given in terms of the minimum information
$I_{\min}$. Unlike the I-measure \cite{yeung2012first}, each atom
in \cite{williams2010nonnegative} has a nonnegative measure. The
partial information decomposition has been further studied in \cite{harder2013bivariate,bertschinger2013shared,griffith2014quantifying,olbrich2015information,rosas2020operational},
and various measures other than the minimum information have been
proposed. In this paper, we can write the minimum information in the
original partial information decomposition work \cite{williams2010nonnegative}
in terms of the Poisson information measure in \eqref{eq:min_info}.

The analogues of Boolean set operations between random variables,
where the results of the operations must also be random variables,
have been systematically studied in \cite{li2017extended}. It was
noted that the only set operation that can be defined unambiguously
is the union (corresponding to joint random variable). Operations
such as intersection and difference correspond to trade-off regions,
and will result in different random variables depending on which quanitities
are minimized/maximized. In \cite{ellerman2017logical,down2023logarithmic}
and this work, we relax the domain to be a ``generalized information
space'' that can contain more than only random variables, allowing
the set operations to be defined canonically.

Another approach to decompose information has been studied in \cite{li2022infinite},
which showed that information is ``approximately infinitely divisible'',
i.e., a random variable can be written as a function of arbitrarily
many i.i.d. random variables, with a multiplicative gap on the entropy.
Nevertheless, this decomposition is not suitable for the purpose of
this paper, since the decomposition of $X$ and the decomposition
of $Y$ do not generally contain common random variables even when
$I(X;Y)\neq0$, and hence set intersection between the decompositions
does not correspond to the mutual information.

The homological nature of entropy and (multivariate) mutual information
has been studied in \cite{baudot2015homological}, which provides
insights into the relations between entropy and mutual information.
The information diagram and its generalizations to other measures
has been studied in \cite{lang2022information}. 

\medskip{}

\subsection*{Notations}

We assume all probability spaces in this paper are over the Borel
sigma-algebras of Polish spaces. Write $[n]=\{1,\ldots,n\}$. Random
variables are usually denoted as $X,Y,Z$. Events are usually denoted
as $E,F$. For an event or a statement $E$, the indicator $\mathbf{1}\{E\}\in\{0,1\}$
is $1$ if $E$ occurs, or $0$ if $E$ does not occur. For a discrete
random variable $X\in\mathcal{X}$, its self-information at $x\in\mathcal{X}$
is $\iota_{X}(x)=-\log p_{X}(x)$. For two jointly-distributed discrete
random variables $X,Y$, their joint self-information at $(x,y)$
is $\iota_{X,Y}(x,y)=-\log p_{X,Y}(x,y)$, and their information density
at $(x,y)$ is $\iota_{X;Y}(x;y)=\iota_{X}(x)+\iota_{Y}(y)-\iota_{X,Y}(x,y)$. 

\medskip{}

\section{Poisson Information Measure}

Recall that the Shannon entropy $H(X)$ of a random variable is given
by the expectation of the self-information $\mathbb{E}[\iota_{X}(X)]$.
The self-information can be regarded as a ``pointwise entropy''
which refines upon the entropy. In this section, we introduce the
Poisson information measure of a random variable, which can be regarded
as a ``further refinement'' of the self-information.
\begin{defn}
\label{def:poient_rv}For $u\in\Omega$ and $\mathbf{v}=\{(v_{i},t_{i}):i\in\mathbb{N}\}$,
$v_{i}\in\Omega$, $0<t_{1}\le t_{2}\le\cdots$, and $X$ being a
discrete random variable over the probability space $(\Omega,\mathcal{F},\mathbb{P})$,
the \emph{Poisson information measure} of $X$ at $u$ with respect
to $\mathbf{v}$ is
\[
\mathcal{H}_{u;\mathbf{v}}(X)=\log\frac{\min_{i\in\mathbb{N}:\,X(v_{i})=X(u)}t_{i}}{\min_{i\in\mathbb{N}}t_{i}}.
\]
Let $\mathcal{H}_{u;\mathbf{v}}(X)=\infty$ if there does not exist
$i$ such that $X(v_{i})=X(u)$.
\end{defn}
\medskip{}

We can regard $v_{i}$ as the position of the $i$-th point, and $t_{i}$
as the arrival time of the $i$-th point. The Poisson information
measure is the logarithm of the first time we encounter a point that
gives the same value of $X$ as $u$, divided by the first time any
point arrives. Intuitively, the rarer the value $X(u)$ is, the more
points we need to scan until we find $X(v_{i})=X(u)$, and the larger
$\mathcal{H}_{u;\mathbf{v}}(X)$ will be.

We say that $\mathbf{V}=\{(V_{i},T_{i}):i\in\mathbb{N}\}$ is a $\mathbb{P}$\emph{-labeled
Poisson process with rate $\lambda$} if $V_{i}\sim\mathbb{P}$ i.i.d.,
independent of $(T_{i})_{i\in\mathbb{N}}$ which form a Poisson process
with rate $\lambda$, i.e., $T_{1}$, $T_{2}-T_{1}$, $T_{3}-T_{2},\ldots$
are i.i.d. $\mathrm{Exp}(\lambda)$ random variables. By taking $\mathbf{V}$
to be a $\mathbb{P}$-labeled Poisson process with rate $1$ and taking
expectation, we can obtain the average Poisson entropy.

\medskip{}

\begin{defn}
\label{def:poient_rv_avg}For $u\in\Omega$ and $X$ being a discrete
random variable, the \emph{(average) Poisson information measure}
of $X$ at $u$ is
\[
\mathcal{H}_{u}(X)=\mathbb{E}\left[\mathcal{H}_{u;\mathbf{V}}(X)\right],
\]
where $\mathbf{V}$ is a $\mathbb{P}$-labeled Poisson process with
rate $1$. We further define the \emph{(average) Poisson information
measure} of $X$ as
\[
\mathcal{H}(X)=\mathbb{E}\left[\mathcal{H}_{U;\mathbf{V}}(X)\right],
\]
where $U\sim\mathbb{P}$ is independent of the $\mathbb{P}$-labeled
Poisson process $\mathbf{V}$.
\end{defn}
\medskip{}

It is straightforward to check that these notions of average Poisson
information measure are simply the familiar notions of self-information
and Shannon entropy.

\medskip{}

\begin{prop}
\label{prop:poient_ent}We have
\[
\mathcal{H}_{u}(X)=\iota_{X}(X(u)),
\]
\[
\mathcal{H}(X)=H(X).
\]
\end{prop}
\begin{IEEEproof}
We have
\begin{align*}
\mathcal{H}_{u}(X) & =\mathbb{E}\left[\mathcal{H}_{u;\mathbf{V}}(X)\right]\\
 & =\mathbb{E}\left[\log\frac{\min_{i\in\mathbb{N}:\,X(V_{i})=X(u)}t_{i}}{\min_{i\in\mathbb{N}}t_{i}}\right]\\
 & =\mathbb{E}\left[\log\Big(\mathbb{P}(X^{-1}(X(u)))\min_{i\in\mathbb{N}:\,X(V_{i})=X(u)}t_{i}\Big)\right]-\mathbb{E}\left[\log\min_{i\in\mathbb{N}}t_{i}\right]+\iota_{X}(X(u))\\
 & =\iota_{X}(X(u)),
\end{align*}
where the last line is because $\mathbb{P}(X^{-1}(X(u)))\min_{i\in\mathbb{N}:\,X(V_{i})=X(u)}t_{i}$
and $\min_{i\in\mathbb{N}}t_{i}$ are both $\mathrm{Exp}(1)$ random
variables by the thinning property of Poisson process \cite{kingman1992poisson}.
Hence we also have $\mathcal{H}(X)=\mathbb{E}[\mathcal{H}_{U}(X)]=\mathbb{E}[\iota_{X}(X(U))]=H(X)$.
\end{IEEEproof}
\medskip{}

\section{Generalized Information Space}

In this section, we will define the mapping from random variables
to sets. We first define the space in which these sets reside.

\medskip{}

\begin{defn}
\label{def:geninfospace}Given a sigma algebra $\mathcal{F}$ over
the sample space $\Omega$, its \emph{generalized information space}
is defined as 
\[
\mathrm{G}(\mathcal{F}):=\bigsqcup_{k=1}^{\infty}\mathcal{F}^{\otimes k},
\]
where $\mathcal{F}^{\otimes k}$ denotes the $k$-th tensor power
sigma algebra, and $\bigsqcup_{k=1}^{\infty}\mathcal{F}^{\otimes k}$
denotes the sigma algebra generated by the disjoint union of $\mathcal{F}^{\otimes k}$.
More explicitly, $\mathrm{G}(\mathcal{F})$ is the sigma algebra generated
by the sets
\[
\left\{ S_{1}\times\cdots\times S_{k}:\,k\ge1,\,S_{1},\ldots,S_{k}\in\mathcal{F}\right\} .
\]
A \emph{generalized information} over $\mathcal{F}$ is an element
of $\mathrm{G}(\mathcal{F})$. 
\end{defn}
\medskip{}

Note that a generalized information is a set of tuples in the form
$(\omega_{1},\ldots,\omega_{k})$ (with any positive length) where
$\omega_{i}\in\Omega$. If $\mathcal{F}=2^{\Omega}$ is the set of
all subsets of $\Omega$ (e.g. for a finite $\Omega$), then $\mathrm{G}(\mathcal{F})$
is simply given by $2^{\bigcup_{k=1}^{\infty}\Omega^{k}}$. 

Before we define the mapping from a random variable to a generalized
information, we first define the mapping for events.

\medskip{}

\begin{defn}
\label{def:gen_event}The \emph{associated generalized information}
of an event $E\in\mathcal{F}$ is
\[
\mathrm{G}(E)=\bigcup_{k=1}^{\infty}E^{k},
\]
i.e., $\mathrm{G}(E)$ consists of all tuples $(\omega_{1},\ldots,\omega_{k})$
where $\omega_{1},\ldots,\omega_{k}\in E$.
\end{defn}
\medskip{}

Note that for two events $E_{1},E_{2}$, we have
\begin{equation}
\mathrm{G}(E_{1}\cap E_{2})=\mathrm{G}(E_{1})\cap\mathrm{G}(E_{2}).\label{eq:gen_int}
\end{equation}
Nevertheless, for union, we only have $\mathrm{G}(E_{1})\cup\mathrm{G}(E_{2})\subseteq\mathrm{G}(E_{1}\cup E_{2})$,
and the two sides are generally not equal.

We can then define the generalized information associated with a random
variable.

\medskip{}

\begin{defn}
\label{def:gen_rv}The \emph{associated generalized information} of
a random variable $X\in\mathcal{X}$ is
\[
\tilde{\mathrm{G}}(X)=\mathrm{G}(\Omega)\,\backslash\,\bigcup_{x\in\mathcal{X}}\mathrm{G}(X^{-1}(x)),
\]
where $X^{-1}(x)$ is the event ``$X=x$''.
\end{defn}
\medskip{}

\begin{figure}
\begin{centering}
\includegraphics[scale=0.78]{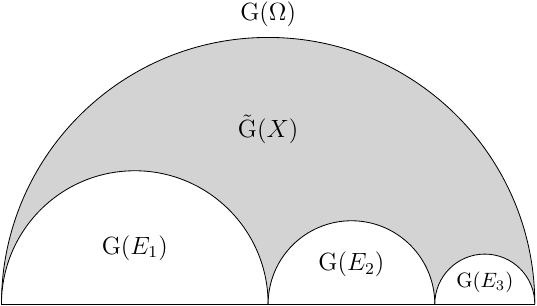}
\par\end{centering}
\caption{\label{fig:rv}Let $X$ be a random variable that induces the partition
$\{E_{1},E_{2},E_{3}\}$ of $\Omega$. Then $\tilde{\mathrm{G}}(X)$
(shaded area), $\mathrm{G}(E_{1}),\mathrm{G}(E_{2}),\mathrm{G}(E_{3})$
is a partition of $\mathrm{G}(\Omega)$ (the whole semicircle).}
\end{figure}

In other words, $\tilde{\mathrm{G}}(X)$ consists of all tuples $(\omega_{1},\ldots,\omega_{k})$
where $X(\omega_{1}),\ldots,X(\omega_{k})$ are not all equal.\footnote{We have $\mathrm{G}(X)\in\mathrm{G}(\mathcal{F})$ since diagonal
sets of the tensor power of the Borel sigma algebra of a Polish space
is measurable.} An elegant consequence is that the following sets
\[
\tilde{\mathrm{G}}(X),\,\mathrm{and}\;\mathrm{G}(X^{-1}(x))\;\mathrm{for}\;x\in\mathcal{X}
\]
form a partition of $\mathrm{G}(\Omega)$. See Figure \ref{fig:rv}
for an illustration. 

Note that although $X^{-1}(x)$ for $x\in\mathcal{X}$ form a partition
of $\Omega$, $\mathrm{G}(X^{-1}(x))$ for $x\in\mathcal{X}$ are
disjoint but does not form a partition of $\mathrm{G}(\Omega)$ in
general. The generalized information $\tilde{\mathrm{G}}(X)$ is the
remaining part of $\mathrm{G}(\Omega)$ not covered by $\mathrm{G}(X^{-1}(x))$
for $x\in\mathcal{X}$, which can be regarded as the ``information
that is not localized within $X^{-1}(x)$ for any $x$''.

Definition \ref{def:gen_rv} shares some similarities with \cite{down2023logarithmic},
which maps a random variable $X$ to the set $\mathcal{C}(X)=\{B\in2^{\Omega}:\,|X(B)|>1\}\in2^{2^{\Omega}}$,
which is a set of subsets of $\Omega$. Note that Definition \ref{def:gen_rv}
can be written as $\tilde{\mathrm{G}}(X)=\{(\omega_{1},\ldots,\omega_{k}):\,|X(\{\omega_{1},\ldots,\omega_{k}\})|>1\}$.
A major difference is that Definition \ref{def:gen_rv} maps a random
variable to a set of tuples with entries in $\Omega$. Unlike a set,
entries in a tuple are not interchangeable. We will see later in Section
\ref{sec:crossent} that the order of the entries in the tuple (or
at least which entry is the first) is essential to obtaining the cross
entropy and the Kullback-Leibler divergence.

For a random variable $X$, and for $U_{1},\ldots,U_{k}\stackrel{iid}{\sim}\mathbb{P}$
($k\ge2$), we have
\begin{align*}
\mathbb{P}\big((U_{1},\ldots,U_{k})\in\tilde{\mathrm{G}}(X)\big) & =1-\sum_{x\in\mathcal{X}}(p_{X}(x))^{k}\\
 & =(k-1)S_{k}(X),
\end{align*}
where $S_{k}(X)$ is the order-$k$ Tsallis entropy \cite{tsallis1988possible}.
The case $k=2$ of this fact has been used in the construction in
\cite{ellerman2017logical}. Although the order-$1$ Tsallis entropy
is the Shannon entropy, the above equation is meaningless for $k=1$,
and more work is needed to define a measure that recovers the Shannon
entropy. 

We will then recover the Shannon entropy by showing that the Poisson
information measure $\mathcal{H}_{u;\mathbf{v}}$ applies not only
to random variables, but can be generalized to be a ``signed measure''
over $\mathrm{G}(\mathcal{F})$.

\medskip{}

\begin{defn}
\label{def:poient_gen}For a generalized information $A\in\mathrm{G}(\mathcal{F})$,
$u\in\Omega$ and $\mathbf{v}=\{(v_{i},t_{i}):i\in\mathbb{N}\}$,
$v_{i}\in\Omega$, $0<t_{1}\le t_{2}\le\cdots$, the \emph{Poisson
information measure} of $A$ at $u$ with respect to $\mathbf{v}$
is
\[
\mathcal{H}_{u;\mathbf{v}}(A)=\sum_{i=1}^{\infty}\Big(\log\frac{t_{i+1}}{t_{i}}\Big)\sum_{S\subseteq[i]}(-1)^{|S|+1}\mathbf{1}\big\{(u,v_{S_{(1)}},\ldots,v_{S_{(|S|)}})\in A\big\}
\]
where we assume $S=\{S_{(1)},\ldots,S_{(|S|)}\}$, $S_{(1)}<\cdots<S_{(|S|)}$.
The \emph{(average) Poisson information measures} are again defined
as
\[
\mathcal{H}_{u}(A)=\mathbb{E}\left[\mathcal{H}_{u;\mathbf{V}}(A)\right],
\]
where $\mathbf{V}$ is a $\mathbb{P}$-labeled Poisson process with
rate $1$, and
\[
\mathcal{H}(A)=\mathbb{E}\left[\mathcal{H}_{U;\mathbf{V}}(A)\right],
\]
where $U\sim\mathbb{P}$ is independent of the $\mathbb{P}$-labeled
Poisson process $\mathbf{V}$.
\end{defn}
\medskip{}

 It is intuitively obvious from Definition \ref{def:poient_gen}
that $\mathcal{H}_{u;\mathbf{v}}$, $\mathcal{H}_{u}$ and $\mathcal{H}$
are (loosely speaking) signed measures over $\mathrm{G}(\Omega)$,
where the total measure is
\[
\mathcal{H}_{u;\mathbf{v}}(\mathrm{G}(\Omega))=\mathcal{H}_{u}(\mathrm{G}(\Omega))=\mathcal{H}(\mathrm{G}(\Omega))=0.
\]
Nevertheless, the infinite sum in Definition \ref{def:poient_gen}
might not converge in general. Therefore, we will only prove the following
weaker result that shows the average Poisson information measure $\mathcal{H}$
is a finitely-additive signed measure. The proof is given in Appendix
\ref{subsec:pf_finadd_meas}.

\medskip{}

\begin{prop}
\label{prop:finadd_meas}Consider $\mathrm{G}_{\mathrm{fin}}(\mathcal{F})$,
defined as the field of sets generated by $\{\mathrm{G}(E):\,E\in\mathcal{F}\}$.\footnote{A field of sets over the set $S$ is a set $\mathcal{G}\subseteq2^{S}$
that is closed under complement, finite union and finite intersection,
and contains $\emptyset$. The field of sets generated by $\mathcal{T}\subseteq2^{S}$
is the smallest field of sets $\mathcal{G}$ satisfying that $\mathcal{T}\subseteq\mathcal{G}$
(it is the closure of $\mathcal{T}$ under complement, finite union
and finite intersection, and contains $\emptyset$). A finitely-additive
signed measure (or a finitely-additive set function) over a field
of sets $\mathcal{G}$ is a function $\mu:\mathcal{G}\to\mathbb{R}$
such that $\mu(\emptyset)=0$ and $\mu(A\cup B)=\mu(A)+\mu(B)$ for
disjoint $A,B\in\mathcal{G}$.} Then $\mathcal{H}$ is a finitely-additive signed measure over $\mathrm{G}_{\mathrm{fin}}(\mathcal{F})$.
\end{prop}
\medskip{}

To check that Definition \ref{def:poient_gen} is a generalization
of Definition \ref{def:poient_rv}, we have to show that they coincide
for random variables.

\medskip{}

\begin{prop}
For a discrete random variable $X$,
\[
\mathcal{H}_{u;\mathbf{v}}(\tilde{\mathrm{G}}(X))=\mathcal{H}_{u;\mathbf{v}}(X).
\]
Hence,
\[
\mathcal{H}_{u}(\tilde{\mathrm{G}}(X))=\iota_{X}(X(u)),
\]
\[
\mathcal{H}(\tilde{\mathrm{G}}(X))=H(X).
\]
\end{prop}
\begin{IEEEproof}
We have
\begin{align*}
\mathcal{H}_{u;\mathbf{v}}(X) & =\log\frac{\min_{i\in\mathbb{N}:\,X(v_{i})=X(u)}t_{i}}{\min_{i\in\mathbb{N}}t_{i}}\\
 & =\sum_{i=1}^{\min\{j:\,X(v_{j})=X(u)\}-1}\log\frac{t_{i+1}}{t_{i}}\\
 & =\sum_{i=1}^{\infty}\Big(\log\frac{t_{i+1}}{t_{i}}\Big)\mathbf{1}\big\{\forall j\in[i]:\,X(v_{j})\neq X(u)\big\}\\
 & \stackrel{(a)}{=}\sum_{i=1}^{\infty}\Big(\log\frac{t_{i+1}}{t_{i}}\Big)\sum_{S\subseteq[i]}(-1)^{|S|+1}\mathbf{1}\big\{\exists j\in S:\,X(v_{j})\neq X(u)\big\}\\
 & \stackrel{(b)}{=}\sum_{i=1}^{\infty}\Big(\log\frac{t_{i+1}}{t_{i}}\Big)\sum_{S\subseteq[i]}(-1)^{|S|+1}\mathbf{1}\big\{(u,v_{S_{(1)}},\ldots,v_{S_{(|S|)}})\in\tilde{\mathrm{G}}(X)\big\},
\end{align*}
where (a) is by the inclusion-exclusion principle, and (b) is because
$(u,v_{S_{(1)}},\ldots,v_{S_{(|S|)}})\in\mathrm{G}(X)$ if and only
if $X(u),X(v_{S_{(1)}}),\ldots,X(v_{S_{(|S|)}})$ are not all equal.
As a result, by Proposition \ref{prop:poient_ent}, $\mathcal{H}_{u}(\tilde{\mathrm{G}}(X))=\iota_{X}(X(u))$
and $\mathcal{H}(\tilde{\mathrm{G}}(X))=H(X)$.
\end{IEEEproof}
\medskip{}

We can also evaluate $\mathcal{H}_{u;\mathbf{v}}(\mathrm{G}(E))$
for event $E\in\mathcal{F}$.

\medskip{}

\begin{prop}
\label{prop:poimeas_event}For an event $E\in\mathcal{F}$,
\[
\mathcal{H}_{u;\mathbf{v}}(\mathrm{G}(E))=-\mathbf{1}\{u\in E\}\log\frac{\min_{i\in\mathbb{N}:\,v_{i}\in E}t_{i}}{\min_{i\in\mathbb{N}}t_{i}}.
\]
Hence, 
\[
\mathcal{H}_{u}(\mathrm{G}(E))=\mathbf{1}\{u\in E\}\log\mathbb{P}(E),
\]
\[
\mathcal{H}(\mathrm{G}(E))=\mathbb{P}(E)\log\mathbb{P}(E).
\]
\end{prop}
\begin{IEEEproof}
We have 
\begin{align*}
 & \mathcal{H}_{u;\mathbf{v}}(\mathrm{G}(E))\\
 & =\sum_{i=1}^{\infty}\Big(\log\frac{t_{i+1}}{t_{i}}\Big)\sum_{S\subseteq[i]}(-1)^{|S|+1}\mathbf{1}\big\{(u,v_{S_{(1)}},\ldots,v_{S_{(|S|)}})\in\mathrm{G}(E)\big\}\\
 & =\mathbf{1}\{u\in E\}\sum_{i=1}^{\infty}\Big(\log\frac{t_{i+1}}{t_{i}}\Big)\sum_{S\subseteq\{j\in[i]:\,v_{j}\in E\}}(-1)^{|S|+1}\\
 & =-\mathbf{1}\{u\in E\}\sum_{i=1}^{\infty}\Big(\log\frac{t_{i+1}}{t_{i}}\Big)\cdot\mathbf{1}\left\{ \{j\in[i]:\,v_{j}\in E\}=\emptyset\right\} \\
 & =-\log\frac{\min_{i\in\mathbb{N}:\,v_{i}\in E}t_{i}}{\min_{i\in\mathbb{N}}t_{i}}.
\end{align*}
\end{IEEEproof}
\medskip{}

Note that the Poisson information measure is positive for random variables,
and negative for events. Since $\tilde{\mathrm{G}}(X)$ and $\mathrm{G}(X^{-1}(x))$
for $x\in\mathcal{X}$ partition $\mathrm{G}(\Omega)$, the positive
term and the negative terms cancel out in $\mathcal{H}(\tilde{\mathrm{G}}(X))+\sum_{x}\mathcal{H}(\mathrm{G}(X^{-1}(x)))=\mathcal{H}(\mathrm{G}(\Omega))=0$.
We choose the sign this way so the Poisson information measure for
random variables coincides with the Shannon entropy. Nevertheless,
it can be argued that we should define it the other way around (negative
for random variables, positive for events) since information reduces
uncertainty, and hence has a net negative effect on the entropy. Ultimately,
the choice of sign in the definition is rather arbitrary.

We now show that Definition \ref{def:poient_gen}, when applied on
a finite partition, recovers the logarithmic decomposition for information
in \cite{down2023logarithmic}.  Consider disjoint events $B_{1},\ldots,B_{n}\in\mathcal{F}$,
and define 
\begin{align*}
 & \mathrm{G}(B_{1};\ldots;B_{n})\\
 & =\Big\{(\omega_{1},\ldots,\omega_{k})\in\mathrm{G}(\Omega):\\
 & \;\;\;\;\;\{\{i\in[k]:\,\omega_{i}\in B_{j}\}:\,j\in[n]\}\;\text{is a partition of}\;[k]\Big\}\\
 & =\mathrm{G}\big(\bigcup_{j\in[n]}B_{j}\big)\backslash\bigcup_{i\in[n]}\mathrm{G}\big(\bigcup_{j\in[n]\backslash\{i\}}B_{j}\big).
\end{align*}
In other words, $\mathrm{G}(B_{1};\ldots;B_{n})$ contains each tuple
that has at least one entry in $B_{j}$ for each $j$, and nothing
outside of $\bigcup_{j}B_{j}$. Note that this recovers Definition
\ref{def:gen_event} when $n=1$.

We now show that the Poisson information measure of $\mathrm{G}(B_{1};\ldots;B_{n})$
recovers the formula of interior loss in \cite[Lemma 8]{down2023logarithmic}.

\medskip{}

\begin{prop}
For disjoint events $B_{1},\ldots,B_{n}\in\mathcal{F}$, letting $p(S)=\sum_{i\in S}\mathbb{P}(B_{i})$
for $S\subseteq[n]$, we have
\[
\mathcal{H}(\mathrm{G}(B_{1};\ldots;B_{n}))=\sum_{S\subseteq[n]}(-1)^{|S|+n}p(S)\log p(S).
\]
\end{prop}
\begin{IEEEproof}
By the inclusion-exclusion principle,
\begin{align*}
 & \mathcal{H}\bigg(\bigcup_{i\in[n]}\mathrm{G}\big(\bigcup_{j\in[n]\backslash\{i\}}B_{j}\big)\bigg)\\
 & =\sum_{S\subseteq[n],\,S\neq\emptyset}(-1)^{|S|+1}\mathcal{H}\bigg(\bigcap_{i\in S}\mathrm{G}\big(\bigcup_{j\in[n]\backslash\{i\}}B_{j}\big)\bigg)\\
 & \stackrel{(a)}{=}\sum_{S\subseteq[n],\,S\neq\emptyset}(-1)^{|S|+1}\mathcal{H}\bigg(\mathrm{G}\big(\bigcup_{j\in[n]\backslash S}B_{j}\big)\bigg)\\
 & \stackrel{(b)}{=}\sum_{S\subseteq[n],\,S\neq\emptyset}(-1)^{|S|+1}p([n]\backslash S)\log p([n]\backslash S)\\
 & =\sum_{S\subsetneq[n]}(-1)^{|S|+n+1}p(S)\log p(S),
\end{align*}
where (a) is by \eqref{eq:gen_int}, and (b) is by Proposition \eqref{prop:poimeas_event}.
Hence
\begin{align*}
 & \mathcal{H}(\mathrm{G}(B_{1};\ldots;B_{n}))\\
 & =\mathcal{H}\bigg(\mathrm{G}\big(\bigcup_{j\in[n]}B_{j}\big)\backslash\bigcup_{i\in[n]}\mathrm{G}\big(\bigcup_{j\in[n]\backslash\{i\}}B_{j}\big)\bigg)\\
 & =\sum_{S\subseteq[n]}(-1)^{|S|+n}p(S)\log p(S).
\end{align*}
\end{IEEEproof}
\medskip{}

Therefore, the Poisson information measure can be regarded as a generalization
of \cite{down2023logarithmic} (which only concerns finite probability
spaces) to general probability spaces. For a general probability space,
the generalized information $\tilde{\mathrm{G}}(X)$ of all random
variables $X$ can be defined over the same space of generalized information
$\mathrm{G}(\mathcal{F})$. Although Proposition \ref{prop:finadd_meas}
only shows that $\mathcal{H}$ is a finitely-additive signed measure,
it is still strictly stronger than \cite{down2023logarithmic} since
$\mathcal{H}$ here is defined over a field of sets $\mathrm{G}_{\mathrm{fin}}(\mathcal{F})$
which can be infinite, over which infinitely many discrete random
variables can be defined. Note that although this paper allows a continuous
probability space, for a continuous random variable $X$, $\mathcal{H}(\tilde{\mathrm{G}}(X))$
does not give the differential entropy of $X$, but instead gives
$\mathcal{H}(\tilde{\mathrm{G}}(X))=\infty$, which is the correct
value of the (discrete) Shannon entropy of a continuous random variable.\footnote{A common misconception is that there is a single notion of ``entropy''
which takes a distribution as input, and outputs the discrete Shannon
entropy if the distribution is discrete, and the differential entropy
if the distribution is continuous. This is false, for the discrete
entropy and the differential entropy are two very different concepts
(e.g. discrete entropy is invariant under bijective mappings, but
differential entropy is not). The correct value of the discrete entropy
of a continuous random variable is $\infty$, whereas the correct
value of the differential entropy of a discrete random variable is
$-\infty$. }

\medskip{}

\section{Joint Random Variables as Set Unions}

Similar to \cite{ellerman2017logical,down2023logarithmic}, the joint
random variable $(X,Y)$ corresponds to set union $\tilde{\mathrm{G}}(X)\cup\tilde{\mathrm{G}}(Y)$
in the space of generalized information. 
\begin{prop}
\label{prop:union}For finite discrete random variables $X,Y$, we
have 
\[
\tilde{\mathrm{G}}((X,Y))=\tilde{\mathrm{G}}(X)\cup\tilde{\mathrm{G}}(Y).
\]
Hence,
\[
\mathcal{H}_{u;\mathbf{v}}(\tilde{\mathrm{G}}(X)\cup\tilde{\mathrm{G}}(Y))=\log\frac{\min_{i\in\mathbb{N}:\,X(v_{i})=X(u),\,Y(v_{i})=Y(u)}t_{i}}{\min_{i\in\mathbb{N}}t_{i}},
\]
\[
\mathcal{H}_{u}(\tilde{\mathrm{G}}(X)\cup\tilde{\mathrm{G}}(Y))=\iota_{X,Y}(X(u),Y(u)),
\]
\[
\mathcal{H}(\tilde{\mathrm{G}}(X)\cup\tilde{\mathrm{G}}(Y))=H(X,Y).
\]
\end{prop}
\begin{IEEEproof}
We have
\begin{align*}
\tilde{\mathrm{G}}((X,Y)) & =\mathrm{G}(\Omega)\,\backslash\,\bigcup_{(x,y)\in\mathcal{X}\times\mathcal{Y}}\mathrm{G}(X^{-1}(x)\cap Y^{-1}(y))\\
 & =\mathrm{G}(\Omega)\,\backslash\,\bigcup_{(x,y)\in\mathcal{X}\times\mathcal{Y}}\big(\mathrm{G}(X^{-1}(x))\cap\mathrm{G}(Y^{-1}(y))\big)\\
 & =\mathrm{G}(\Omega)\,\backslash\,\Big(\bigcup_{x\in\mathcal{X}}\mathrm{G}(X^{-1}(x))\Big)\cap\Big(\bigcup_{y\in\mathcal{Y}}\mathrm{G}(Y^{-1}(y))\Big)\\
 & =\Big(\mathrm{G}(\Omega)\,\backslash\,\bigcup_{x\in\mathcal{X}}\mathrm{G}(X^{-1}(x))\Big)\cup\Big(\mathrm{G}(\Omega)\,\backslash\,\bigcup_{y\in\mathcal{Y}}\mathrm{G}(Y^{-1}(y))\Big).
\end{align*}
\end{IEEEproof}
\medskip{}

\section{Conditioning on Events as Set Intersection}

Given an event $E\in\mathcal{F}$, a \emph{conditional random variable}
given $E$ is a random variable defined over the restriction of $\mathcal{F}$
to $E$, i.e., it is a measurable function $X:E\to\mathcal{X}$. Intuitively,
it is a random variable that is only defined over $E$. In this paper,
we will represent a conditional random variable as the generalized
information
\[
\mathrm{G}(E)\,\backslash\,\bigcup_{x\in\mathcal{X}}\mathrm{G}(X^{-1}(x)),
\]
which generalizes Definition \ref{def:gen_rv} (which concerns ordinary
random variables, i.e., conditional random variables given $\Omega$).

For any (ordinary) random variable $X:\Omega\to\mathcal{X}$ and event
$E$, we can form a conditional random variable ``$X|_{E}$'' by
restricting the domain of $X$ to $E$, i.e., $X|_{E}:E\to\mathcal{X}$
is defined by $X|_{E}(\omega)=X(\omega)$. It can be checked that
the generalized information representing $X|_{E}$ is given by
\begin{align*}
 & \mathrm{G}(E)\,\backslash\,\bigcup_{x\in\mathcal{X}}\mathrm{G}(X|_{E}^{-1}(x))\\
 & =\tilde{\mathrm{G}}(X)\cap\mathrm{G}(E),
\end{align*}
i.e., the operation of conditioning a random variable on an event
is simply given by intersection in the domain of generalized information.
Intuitively, intersecting with $\mathrm{G}(E)$ restricts the information
of interest to $E$, and discards all information outside of $E$.

Recall that the Poisson information measure $\mathcal{H}_{u;\mathbf{v}}$
involves a set of points $\mathbf{v}=\{(v_{i},t_{i}):i\in\mathbb{N}\}$,
$v_{i}\in\Omega$, $0<t_{1}\le t_{2}\le\cdots$, which comes from
a Poisson process. Conditioning on an event $E$ is akin to restricting
attention only to outcomes in $E$, i.e., discarding all points $(v_{i},t_{i})$
where $v_{i}\notin E$. This can be done by considering the intersection
$\mathbf{v}\cap(E\times\mathbb{R})$. More precisely, if $\mathbf{V}$
is a $\mathbb{P}$-labeled Poisson process with rate $\lambda$, then
$\mathbf{V}\cap(E\times\mathbb{R})$ is a $\mathbb{P}|_{E}$-labeled
Poisson process with rate $\lambda\mathbb{P}(E)$ by the thinning
property of Poisson process \cite{kingman1992poisson}, where $\mathbb{P}|_{E}$
is the conditional probability distribution conditional on the event
$E$. Therefore, taking the intersection $\mathbf{v}\cap(E\times\mathbb{R})$
corresponds to conditioning on the event.

We then show that intersecting with $\mathrm{G}(E)$ indeed lets us
ignore the points $(v_{i},t_{i})$ where $v_{i}\notin E$, and consider
only the points in $\mathbf{v}\cap(E\times\mathbb{R})$ in the calculation
of $\mathcal{H}_{u;\mathbf{v}}$.

\medskip{}

\begin{prop}
\label{prop:cond_event}For any event $E\in\mathcal{F}$ and generalized
information $A\in\mathrm{G}(\mathcal{F})$, we have 
\[
\mathcal{H}_{u;\mathbf{v}}(A\cap\mathrm{G}(E))=\mathbf{1}\{u\in E\}\mathcal{H}_{u;\mathbf{v}\cap(E\times\mathbb{R})}(A)
\]
for any $u,\mathbf{v}$. Hence, for any discrete random variable $X$,
\[
\mathcal{H}_{u;\mathbf{v}}(\tilde{\mathrm{G}}(X)\cap\mathrm{G}(E))=\mathbf{1}\{u\in E\}\log\frac{\min_{i:\,v_{i}\in E,\,X(v_{i})=X(u)}t_{i}}{\min_{i:\,v_{i}\in E}t_{i}},
\]
\[
\mathcal{H}_{u}(\tilde{\mathrm{G}}(X)\cap\mathrm{G}(E))=\mathbf{1}\{u\in E\}\iota_{X|E}(X(u)),
\]
\[
\mathcal{H}(\tilde{\mathrm{G}}(X)\cap\mathrm{G}(E))=\mathbb{P}(E)H(X|E).
\]
\end{prop}
\medskip{}

\begin{IEEEproof}
Let $J=\{j\in\mathbb{N}:\,v_{j}\in E\}$. For $u\in E$, we have
\begin{align*}
 & \mathcal{H}_{u;\mathbf{v}}(A\cap\mathrm{G}(E))\\
 & =\sum_{i=1}^{\infty}\Big(\log\frac{t_{i+1}}{t_{i}}\Big)\sum_{S\subseteq[i]}(-1)^{|S|+1}\mathbf{1}\big\{(u,v_{S_{(1)}},\ldots,v_{S_{(|S|)}})\in A\cap\mathrm{G}(E)\big\}\\
 & =\sum_{i=1}^{\infty}\Big(\log\frac{t_{i+1}}{t_{i}}\Big)\sum_{S\subseteq[i]\cap J}(-1)^{|S|+1}\mathbf{1}\big\{(u,v_{S_{(1)}},\ldots,v_{S_{(|S|)}})\in A\big\}\\
 & \stackrel{(a)}{=}\sum_{i\in J}\Big(\log\frac{t_{\min(J\cap[i+1,\infty))}}{t_{i}}\Big)\sum_{S\subseteq[i]\cap J}(-1)^{|S|+1}\mathbf{1}\big\{(u,v_{S_{(1)}},\ldots,v_{S_{(|S|)}})\in A\big\}\\
 & =\mathcal{H}_{u;\mathbf{v}\cap(E\times\mathbb{R})}(A),
\end{align*}
where (a) is because if $i\notin J$, then $[i]\cap J=[i-1]\cap J$,
and the $i$-th term in the summation can be combined with the $(i-1)$-th
term, and hence we only need to sum over $i\in J$. For $u\notin E$,
we cannot have $(u,v_{S_{(1)}},\ldots,v_{S_{(|S|)}})\in A\cap\mathrm{G}(E)$,
and hence $\mathcal{H}_{u;\mathbf{v}}(A\cap\mathrm{G}(E))=0$. For
the remaining claims, they follow from the fact that if $\mathbf{V}$
is a $\mathbb{P}$-labeled Poisson process with rate $1$, then $\mathbf{V}\cap(E\times\mathbb{R})$
is a $\mathbb{P}|_{E}$-labeled Poisson process with rate $\mathbb{P}(E)$
\cite{kingman1992poisson}, and that the definition of the Poisson
information measure does not depend on the rate of the Poisson process
(the rate is cancelled when we take the ratio $t_{i+1}/t_{i}$).
\end{IEEEproof}
\medskip{}

\section{Conditioning on Random Variables as Set Differences\label{sec:cond_rv}}

\begin{figure}
\begin{centering}
\includegraphics[scale=0.78]{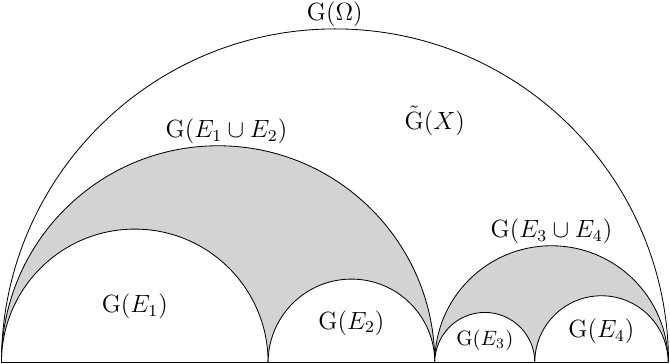}
\par\end{centering}
\caption{\label{fig:rv-1}Let $Y$ be a random variable that induces the partition
$\{E_{1},E_{2},E_{3},E_{4}\}$ of $\Omega$, and $X$ be a random
variable that induces the partition $\{E_{1}\cup E_{2},E_{3}\cup E_{4}\}$
of $\Omega$. The shaded area is $\tilde{\mathrm{G}}(Y)\backslash\tilde{\mathrm{G}}(X)$,
corresponding to ``$Y$ conditional on $X$''.}
\end{figure}

For two generalized information $A,B$, we can interpret the set difference
$A\backslash B$ as ``$A$ conditional on $B$''. In particular,
for two random variables $X,Y$, the generalized information $\mathrm{G}(Y)\backslash\mathrm{G}(X)$
can be written as
\begin{equation}
\tilde{\mathrm{G}}(Y)\backslash\tilde{\mathrm{G}}(X)=\bigcup_{x\in\mathcal{X}}\big(\tilde{\mathrm{G}}(Y)\cap\mathrm{G}(X^{-1}(x))\big)\label{eq:cond_rv}
\end{equation}
by Definition \ref{def:gen_rv}. Note that $\tilde{\mathrm{G}}(Y)\cap\mathrm{G}(X^{-1}(x))$
represents the random variable $Y$ conditional on the event $X^{-1}(x)$.
This gives an intuitive interpretation on ``the information of $Y$
conditional on $X$'' as the union of ``information of $Y$ conditional
on the event $X=x$'' for all values of $x$. See Figure \ref{fig:rv-1}
for an illustration. 

\medskip{}

The following proposition is an immediate consequence of Proposition
\ref{prop:cond_event}. The proof is omitted.

\medskip{}

\begin{prop}
\label{prop:cond_rv}For any discrete random variables $X,Y$,
\[
\mathcal{H}_{u;\mathbf{v}}(\tilde{\mathrm{G}}(Y)\backslash\tilde{\mathrm{G}}(X))=\log\frac{\min_{i\in\mathbb{N}:\,X(v_{i})=X(u),\,Y(v_{i})=Y(u)}t_{i}}{\min_{i\in\mathbb{N}:\,X(v_{i})=X(u)}t_{i}},
\]
\[
\mathcal{H}_{u}(\tilde{\mathrm{G}}(Y)\backslash\tilde{\mathrm{G}}(X))=\iota_{Y|X}(Y(u)|X(u)),
\]
\[
\mathcal{H}(\tilde{\mathrm{G}}(Y)\backslash\tilde{\mathrm{G}}(X))=H(Y|X).
\]
\end{prop}
\medskip{}

\begin{figure}
\begin{centering}
\includegraphics[scale=0.85]{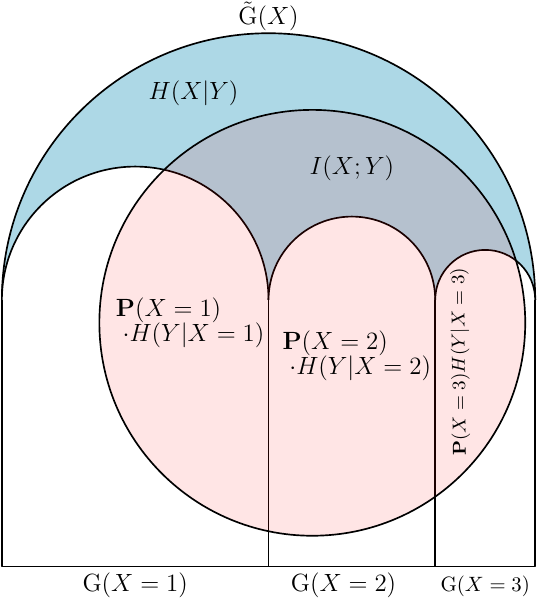}
\par\end{centering}
\caption{\label{fig:condent}Illustration of the identity $H(Y|X)=\sum_{x}p_{X}(x)H(Y|X=x)$.
Let $X\in\{1,2,3\}$ and $Y$ be random variables. Then $\tilde{\mathrm{G}}(X)$
(blue area), $\mathrm{G}(X=1)$, $\mathrm{G}(X=2)$, $\mathrm{G}(X=3)$
form a partition of $\mathrm{G}(\Omega)$ (we write $\mathrm{G}(X=1)=\mathrm{G}(\{X=1\})$
where $\{X=1\}=X^{-1}(1)$ is the event ``$X=1$''). The set $\tilde{\mathrm{G}}(Y)$
(red circle) can be broken into $\tilde{\mathrm{G}}(X)\cap\tilde{\mathrm{G}}(Y)$
with measure $I(X;Y)$, and $\tilde{\mathrm{G}}(Y)\backslash\tilde{\mathrm{G}}(X)$
with measure $H(Y|X)$. The area $\tilde{\mathrm{G}}(Y)\backslash\tilde{\mathrm{G}}(X)$
can be further broken into $\tilde{\mathrm{G}}(Y)\cap\mathrm{G}(X=x)$
(with measure $\mathbb{P}(X=x)H(Y|X=x)$) for $x=1,2,3$.}
\end{figure}

Note that the fact 
\[
H(Y|X)=\sum_{x}p_{X}(x)H(Y|X=x)
\]
is the consequence of taking the measure on both sides of \eqref{eq:cond_rv},
and applying Propositions \ref{prop:cond_event} and \ref{prop:cond_rv}.
This is illustrated in Figure \ref{fig:condent}.

\medskip{}

\section{Mutual Information as Set Intersection}

The conventional insight in \cite{ting1962amount,yeung2012first}
is that the mutual information between two random variables corresponds
to the size of the intersection of their information. Like \cite{down2023logarithmic},
this insight is also realized by the Poisson information measure in
this paper. Moreover, the Poisson information measure can also recover
the information density $\iota_{X;Y}(x;y)$. The following is an immediate
consequence of the inclusion-exclusion principle $\mathcal{H}(\tilde{\mathrm{G}}(X)\cap\tilde{\mathrm{G}}(Y))=\mathcal{H}(\tilde{\mathrm{G}}(X))+\mathcal{H}(\tilde{\mathrm{G}}(Y))-\mathcal{H}(\tilde{\mathrm{G}}(X)\cup\tilde{\mathrm{G}}(Y))$
and Proposition \ref{prop:union}. The proof is omitted.

\medskip{}

\begin{prop}
\label{prop:mutual}For any discrete random variables $X,Y$,
\[
\mathcal{H}_{u;\mathbf{v}}(\tilde{\mathrm{G}}(X)\cap\tilde{\mathrm{G}}(Y))=\log\frac{(\min_{i:\,X(v_{i})=X(u)}t_{i})(\min_{i:\,Y(v_{i})=Y(u)}t_{i})}{(\min_{i}t_{i})(\min_{i:\,X(v_{i})=X(u),\,Y(v_{i})=Y(u)}t_{i})},
\]
\[
\mathcal{H}_{u}(\tilde{\mathrm{G}}(X)\cap\tilde{\mathrm{G}}(Y))=\iota_{X;Y}(X(u);Y(u)),
\]
\[
\mathcal{H}(\tilde{\mathrm{G}}(X)\cap\tilde{\mathrm{G}}(Y))=I(X;Y).
\]
\end{prop}
\medskip{}

Note that we can combine Propositions \ref{prop:cond_rv}, \ref{prop:mutual}
and obtain the conditional mutual information 
\[
\mathcal{H}((\tilde{\mathrm{G}}(X)\cap\tilde{\mathrm{G}}(Y))\backslash\tilde{\mathrm{G}}(Z))=I(X;Y|Z).
\]
The multivariate mutual information $I(X_{1};\cdots;X_{n})$ between
$X_{1},\ldots,X_{n}$ \cite{mcgill1954multivariate} can similarly
be given by $\mathcal{H}(\tilde{\mathrm{G}}(X_{1})\cap\cdots\cap\tilde{\mathrm{G}}(X_{n}))$.

\medskip{}

\section{Cross Entropy and Kullback-Leibler Divergence\label{sec:crossent}}

Recall that Definition \ref{def:gen_rv} maps a random variable to
a set of tuples, unlike \cite{down2023logarithmic} which maps a random
variable to a set of subsets of $\Omega$. Although all constructions
in the previous sections do not concern the order of the entries in
the tuples, we will see in this section that being able to tell the
first entry in a tuple apart from the rest allows us to obtain the
cross entropy and the Kullback-Leibler divergence.

We first define the cross generalized information between two events,
which generalizes Definition \ref{def:gen_event}.

\medskip{}

\begin{defn}
\label{def:gen_event_cross}The \emph{cross generalized information}
from event $F\in\mathcal{F}$ to event $E\in\mathcal{F}$ is
\[
\mathrm{G}(E,F)=\bigcup_{k=0}^{\infty}E\times F^{k},
\]
i.e., $\mathrm{G}(E,F)$ consists of all tuples $(\omega_{1},\ldots,\omega_{k})$
where $\omega_{1}\in E$ and $\omega_{2},\ldots,\omega_{k}\in F$.
\end{defn}
\medskip{}

We list some simple properties of the cross generalized information:
\begin{itemize}
\item $\mathrm{G}(E,E)=\mathrm{G}(E)$.
\item $\mathrm{G}(E_{1},F_{1})\cap\mathrm{G}(E_{2},F_{2})=\mathrm{G}(E_{1}\cap E_{2},F_{1}\cap F_{2})$.
\item $\mathrm{G}(E_{1},F_{1})\cup\mathrm{G}(E_{2},F_{2})\subseteq\mathrm{G}(E_{1}\cup E_{2},F_{1}\cup F_{2})$.
\item $\mathrm{G}(E_{1},F)\cup\mathrm{G}(E_{2},F)=\mathrm{G}(E_{1}\cup E_{2},F)$.
\end{itemize}
The Poisson information measure of $\mathrm{G}(E,F)$ can be given
in a similar manner as Proposition \ref{prop:poimeas_event}. The
proof of the following proposition is omitted.

\medskip{}

\begin{prop}
\label{prop:poimeas_event-1}For events $E,F\in\mathcal{F}$,
\[
\mathcal{H}_{u;\mathbf{v}}(\mathrm{G}(E,F))=-\mathbf{1}\{u\in E\}\log\frac{\min_{i\in\mathbb{N}:\,v_{i}\in F}t_{i}}{\min_{i\in\mathbb{N}}t_{i}},
\]
\[
\mathcal{H}_{u}(\mathrm{G}(E,F))=\mathbf{1}\{u\in E\}\log\mathbb{P}(F),
\]
\[
\mathcal{H}(\mathrm{G}(E,F))=\mathbb{P}(E)\log\mathbb{P}(F).
\]
\end{prop}
\medskip{}

Proposition \ref{prop:cond_event} gives the conditional entropy of
a random variable $X$ given an event $E$ via the intersection $\tilde{\mathrm{G}}(X)\cap\mathrm{G}(E)$.
If we use $\mathrm{G}(E,F)$ instead of $\mathrm{G}(E)$, we obtain
the cross entropy between the conditional distributions instead. The
following is an extension of Proposition \ref{prop:cond_event}. The
proof is similar to Proposition \ref{prop:cond_event} and is omitted.
\begin{prop}
\label{prop:cross_ent}For any events $E,F\in\mathcal{F}$ and generalized
information $A\in\mathrm{G}(\mathcal{F})$, we have 
\[
\mathcal{H}_{u;\mathbf{v}}(A\cap\mathrm{G}(E,F))=\mathbf{1}\{u\in E\}\mathcal{H}_{u;\mathbf{v}\cap(F\times\mathbb{R})}(A)
\]
for any $u,\mathbf{v}$. Hence, for any discrete random variable $X$,
\[
\mathcal{H}_{u;\mathbf{v}}(\tilde{\mathrm{G}}(X)\cap\mathrm{G}(E,F))=\mathbf{1}\{u\in E\}\log\frac{\min_{i:\,v_{i}\in F,\,X(v_{i})=X(u)}t_{i}}{\min_{i:\,v_{i}\in F}t_{i}},
\]
\[
\mathcal{H}_{u}(\tilde{\mathrm{G}}(X)\cap\mathrm{G}(E,F))=\mathbf{1}\{u\in E\}\iota_{X|F}(X(u)),
\]
\[
\mathcal{H}(\tilde{\mathrm{G}}(X)\cap\mathrm{G}(E,F))=\mathbb{P}(E)H(P_{X|E},P_{X|F}),
\]
where $H(P_{X|E},P_{X|F})$ is the cross entropy between the conditional
distributions $P_{X|E},P_{X|F}$.
\end{prop}
\medskip{}

We will now obtain the Kullback-Leibler divergence. First we define
the relative generalized information between two events.

\medskip{}

\begin{defn}
\label{def:gen_event_rel}The \emph{relative generalized information}
from event $F\in\mathcal{F}$ to event $E\in\mathcal{F}$, where $E\subseteq F$,
is
\[
\mathrm{G}(E\Vert F)=\mathrm{G}(E,F)\backslash\mathrm{G}(E),
\]
i.e., $\mathrm{G}(E\Vert F)$ consists of all tuples $(\omega_{1},\ldots,\omega_{k})$
where $\omega_{1}\in E$ and $\omega_{2},\ldots,\omega_{k}\in F$,
but not all of $\omega_{i}$ are in $E$.
\end{defn}
\medskip{}

As a direct consequence of Proposition \ref{prop:cross_ent}, when
$E\subseteq F$, we can obtain the Kullback-Leibler divergence $D_{\mathrm{KL}}(P_{X|E}\,\Vert\,P_{X|F})=H(P_{X|E},P_{X|F})-H(P_{X|E})$
in terms of the measure of the intersection $\tilde{\mathrm{G}}(X)\cap\mathrm{G}(E\Vert F)$.
\begin{prop}
\label{prop:kl}For events $E,F\in\mathcal{F}$, $E\subseteq F$ and
discrete random variable $X$, we have 
\[
\mathcal{H}(\tilde{\mathrm{G}}(X)\cap\mathrm{G}(E\Vert F))=\mathbb{P}(E)D_{\mathrm{KL}}(P_{X|E}\,\Vert\,P_{X|F}).
\]
\end{prop}
\medskip{}

Although the restriction $E\subseteq F$ causes some loss of generality,
it corresponds nicely to the absolute continuity requirement of the
Kullback-Leibler divergence $D_{\mathrm{KL}}(P\Vert Q)$, which is
only defined when $P$ is absolutely continuous with respect to $Q$.
If $E\subseteq F$, $\mathbb{P}(E)>0$, this guarantees that $P_{X|E}$
is absolutely continuous with respect to $P_{X|F}$, and $D_{\mathrm{KL}}(P_{X|E}\,\Vert\,P_{X|F})$
is defined.

\begin{figure}
\begin{centering}
\includegraphics[scale=1.01]{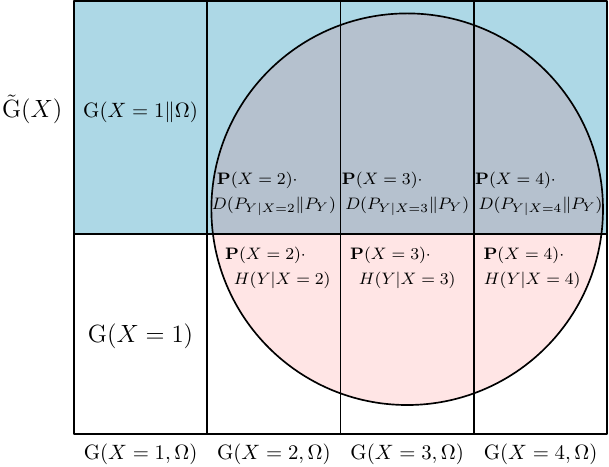}
\par\end{centering}
\caption{\label{fig:condent-1}Illustration of the identity $I(X;Y)=\sum_{x}\mathbb{P}(X=x)D_{\mathrm{KL}}(P_{Y|X=x}\Vert P_{Y})$.
Let $X\in\{1,2,3,4\}$ and $Y$ be random variables. Note that $\mathrm{G}(X=x,\Omega)$
for $x=1,\ldots,4$ partition $\mathrm{G}(\Omega)$. Together with
$\tilde{\mathrm{G}}(X)$ (the blue rectangle), they divide $\mathrm{G}(\Omega)$
into $8$ regions: $\mathrm{G}(X=x\Vert\Omega)$ and $\mathrm{G}(X=x)$
for $x=1,\ldots,4$ (the rectangles in the above figure). The set
$\tilde{\mathrm{G}}(Y)$ (red circle) can be broken into $\tilde{\mathrm{G}}(X)\cap\tilde{\mathrm{G}}(Y)$
with measure $I(X;Y)$, and $\tilde{\mathrm{G}}(Y)\backslash\tilde{\mathrm{G}}(X)$
with measure $H(Y|X)$. As we have seen in Section \ref{sec:cond_rv},
the area $\tilde{\mathrm{G}}(Y)\backslash\tilde{\mathrm{G}}(X)$ can
be further broken into $\tilde{\mathrm{G}}(Y)\cap\mathrm{G}(X=x)$
(with measure $\mathbb{P}(X=x)H(Y|X=x)$) for $x=1,\ldots,4$. This
section shows that the area $\tilde{\mathrm{G}}(X)\cap\tilde{\mathrm{G}}(Y)$
can be further broken into $\tilde{\mathrm{G}}(Y)\cap\mathrm{G}(X=x\Vert\Omega)$
(with measure $\mathbb{P}(X=x)D_{\mathrm{KL}}(P_{Y|X=x}\Vert P_{Y})$)
for $x=1,\ldots,4$.}
\end{figure}

We will now give an intuitive interpretation of the equality

\[
I(X;Y)=\sum_{x\in\mathcal{X}}\mathbb{P}(X=x)D_{\mathrm{KL}}(P_{Y|X=x}\Vert P_{Y}).
\]
It is straightforward to check that for a discrete random variable
$X$, $\tilde{\mathrm{G}}(X)$ can be given as the following disjoint
union
\[
\tilde{\mathrm{G}}(X)=\bigcup_{x\in\mathcal{X}}\mathrm{G}(X=x\Vert\Omega),
\]
where we write $\mathrm{G}(X=x\Vert\Omega)=\mathrm{G}(\{X=x\}\Vert\Omega)$.
Hence we have
\begin{align*}
I(X;Y) & =\mathcal{H}(\tilde{\mathrm{G}}(X)\cap\tilde{\mathrm{G}}(Y))\\
 & =\sum_{x\in\mathcal{X}}\mathcal{H}(\tilde{\mathrm{G}}(Y)\cap\mathrm{G}(X=x\Vert\Omega))\\
 & =\sum_{x\in\mathcal{X}}\mathbb{P}(X=x)D_{\mathrm{KL}}(P_{Y|X=x}\Vert P_{Y}).
\end{align*}
This is illustrated in Figure \ref{fig:condent-1}. Note that the
minimum information $I_{\min}$ in the original partial information
decomposition work \cite{williams2010nonnegative} can then be written
as
\begin{align}
 & I_{\min}(X;\{Y_{1},\ldots,Y_{n}\})\nonumber \\
 & =\sum_{x}\mathbb{P}(X=x)\min_{i}D_{\mathrm{KL}}(P_{Y_{i}|X=x}\Vert P_{Y_{i}})\nonumber \\
 & =\sum_{x}\min_{i}\mathcal{H}(\tilde{\mathrm{G}}(Y_{i})\cap\mathrm{G}(X=x\Vert\Omega)).\label{eq:min_info}
\end{align}

\medskip{}

\section{Information-Event Diagram and the IE-measure}

The \emph{information-event diagram} of the random variables $X_{1},\ldots,X_{n}$
and events $E_{1},\ldots,E_{m}$ is the Venn diagram representing
the sets $\tilde{\mathrm{G}}(X_{1}),\ldots,\tilde{\mathrm{G}}(X_{n}),\mathrm{G}(E_{1}),\ldots,\mathrm{G}(E_{m})$.
This Venn diagram has $2^{n+m}$ cells (note that, unlike the conventional
information diagram, we can also have the cell on the outside $\mathrm{G}(\Omega)\backslash\bigcup_{i}\tilde{\mathrm{G}}(X_{i})\backslash\bigcup_{i}\mathrm{G}(E_{i})$,
which often has a well-defined measure). Although we may also include
the cross generalized information $\mathrm{G}(E_{i},E_{j})$ so we
can discuss the cross entropy and KL divergence, we do not consider
it in this section for the sake of simplicity. 

We can use the following rules to eliminate some cells and simplify
the diagram:
\begin{itemize}
\item \textbf{(A1)}. If $E_{1}\subseteq E_{2}$ are events, then $\mathrm{G}(E_{1})\backslash\mathrm{G}(E_{2})=\emptyset$,
and we can eliminate the cells in the region $\mathrm{G}(E_{1})\backslash\mathrm{G}(E_{2})$.
\item \textbf{(A2)}. If $E_{1}\cap E_{2}=\emptyset$, then $\mathrm{G}(E_{1})\cap\mathrm{G}(E_{2})=\emptyset$.
\item \textbf{(A3)}. If random variable  $Y$ is a function of random variable
$X$ conditional on event $E$ (i.e., $H(Y|X,E)=0$), then $(\tilde{\mathrm{G}}(Y)\backslash\tilde{\mathrm{G}}(X))\cap\mathrm{G}(E)=\emptyset$.
\item \textbf{(A4)}. If random variable $X$ and events $E_{1},\ldots,E_{\ell},F$
satisfy that ``for $u,v\in F$, if $X(u)=X(v)$, then there exists
$E_{i}$ such that $u,v\in E_{i}$'', then 
\[
\mathrm{G}(F)\,\backslash\,\Big(\tilde{\mathrm{G}}(X)\cup\bigcup_{i=1}^{\ell}\mathrm{G}(E_{i})\Big)=\emptyset.
\]
Note that rules A3 and A4 imply that if random variable $X$ induces
a partition $E_{1},\ldots,E_{\ell}$, then $\tilde{\mathrm{G}}(X),\mathrm{G}(E_{1}),\ldots,\mathrm{G}(E_{\ell})$
partition $\mathrm{G}(\Omega)$.
\end{itemize}
Then we can impose the following equalities/inequalities on the measures
of the cells:
\begin{itemize}
\item \textbf{(B1)}. $\mathcal{H}(\mathrm{G}(\Omega))=0$.
\item \textbf{(B2)}. For event $E$, $\mathcal{H}(\mathrm{G}(E))\le0$.
\item \textbf{(B3)}. For random variables $X,Y,Z$ and event $E$, 
\[
\mathcal{H}\Big(\big(\tilde{\mathrm{G}}(X)\cap\tilde{\mathrm{G}}(Y)\cap\mathrm{G}(E)\big)\backslash\tilde{\mathrm{G}}(Z)\Big)\ge0.
\]
This corresponds to $I(X;Y|Z,E)\ge0$.
\end{itemize}
The above rules can also be applied on joint random variables and
intersection of events. For example, if random variables $Y,Z$ are
constant conditional on $E_{1}\cap E_{2}$ (i.e., $Y,Z$ are functions
of the empty random variable $\emptyset$ conditional on $E_{1}\cap E_{2}$),
then $(\tilde{\mathrm{G}}(Y)\cup\tilde{\mathrm{G}}(Z))\cap\mathrm{G}(E_{1})\cap\mathrm{G}(E_{2})=\emptyset$.

\medskip{}

\begin{figure}
\begin{centering}
\includegraphics[scale=1.02]{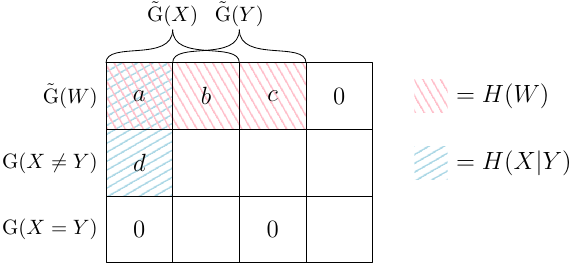}
\par\end{centering}
\caption{\label{fig:fano}The information-event diagram for Fano's inequality.
The three rows are $\tilde{\mathrm{G}}(W),\mathrm{G}(X\protect\neq Y),\mathrm{G}(X=Y)$,
which form a partition of $\mathrm{G}(\Omega)$.}
\end{figure}

\begin{example}
To demonstrate the use of the information-event diagram, we give a
diagrammatic way to present the proof of Fano's inequality \cite{fano1961transmission}:
\begin{equation}
H(X|Y)\le H_{b}(\mathbb{P}(X\neq Y))+\mathbb{P}(X\neq Y)\log(|\mathcal{X}|-1)\label{eq:fano}
\end{equation}
for random variable $X,Y\in\mathcal{X}$, where $H_{b}$ is the binary
entropy function. Let $W=\mathbf{1}\{X\neq Y\}$ be the indicator
random variable. Figure \ref{fig:fano} shows the information-event
diagram of the random variables $X,Y,W$ and the events $\{X\neq Y\},\{X=Y\}$.
Note that $\tilde{\mathrm{G}}(W),\mathrm{G}(X\neq Y),\mathrm{G}(X=Y)$
form a partition of $\mathrm{G}(\Omega)$. Hence the information-event
diagram contains $12$ cells (corresponding to which of $\tilde{\mathrm{G}}(W),\mathrm{G}(X\neq Y),\mathrm{G}(X=Y)$
the cell is in, whether the cell is in $\tilde{\mathrm{G}}(X)$, and
whether the cell is in $\tilde{\mathrm{G}}(Y)$).

There are $3$ cells in Figure \ref{fig:fano} that are $0$, due
to $H(X|Y,\,X=Y)=H(Y|X,\,X=Y)=0$ and $H(W|X,Y)=0$. Let $a,b,c,d$
be the measure of the $4$ cells indicated in Figure \ref{fig:fano}.
We have 
\[
a+b+c=H(W)=H_{b}(\mathbb{P}(X\neq Y)),
\]
\[
b+c=I(Y;W)\ge0,
\]
\[
a+d=H(X|Y),
\]
and
\begin{align*}
d & =\mathbb{P}(X\neq Y)H(X|Y,\,X\neq Y)\\
 & \le\mathbb{P}(X\neq Y)\log(|\mathcal{X}|-1),
\end{align*}
since conditional on $Y=y,\,X\neq Y$, we know $X\in\mathcal{X}\backslash\{y\}$.
Combining these lines gives \eqref{eq:fano}.
\end{example}

\medskip{}

\begin{figure}
\begin{centering}
\includegraphics[scale=1.15]{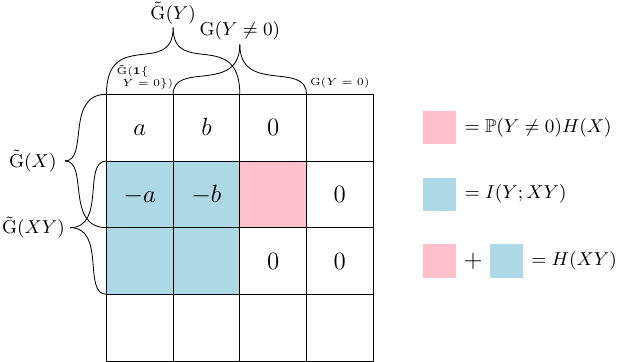}
\par\end{centering}
\caption{\label{fig:mul}Diagrammatic proof of $H(XY)\ge\mathbb{P}(Y\protect\neq0)H(X)$
for independent $X,Y\in\mathbb{R}$.}
\end{figure}

\begin{example}
As a toy example, we prove the claim that for independent real-valued
random variables $X,Y$, we have
\[
H(XY)\ge\mathbb{P}(Y\neq0)H(X).
\]
Consider the information-event diagram with $4$ random variables
$X,Y,XY,\mathbf{1}\{Y=0\}$ and $2$ events $\{Y=0\},\{Y\neq0\}$.
Note that $\tilde{\mathrm{G}}(\mathbf{1}\{Y=0\}),\mathrm{G}(Y=0),\mathrm{G}(Y\neq0)$
form a partition of $\mathrm{G}(\Omega)$. For $\tilde{\mathrm{G}}(Y)$,
it is disjoint of $\mathrm{G}(Y=0)$ due to rule A3, and includes
$\tilde{\mathrm{G}}(\mathbf{1}\{Y=0\})$ also due to rule A3. Hence,
\[
\tilde{\mathrm{G}}(\mathbf{1}\{Y=0\}),\,\mathrm{G}(Y=0),\,\mathrm{G}(Y\neq0)\cap\tilde{\mathrm{G}}(Y),\,\mathrm{G}(Y\neq0)\backslash\tilde{\mathrm{G}}(Y)
\]
form a partition of $\mathrm{G}(\Omega)$. The information-event diagram
can therefore be drawn as Figure \ref{fig:mul}. 

The three cells marked with $0$ are due to $H(XY|Y=0)=0$ and $H(XY|X,Y)=0$.
The two cells marked with $a,-a$ respectively add up to $0$ because
$I(X;\mathbf{1}\{Y=0\})=0$. The two cells marked with $b,-b$ respectively
add up to $0$ because $I(X;Y|\mathbf{1}\{Y=0\})=0$. Therefore, the
red cell in Figure \ref{fig:mul} has a measure 
\begin{align*}
 & \mathcal{H}(\tilde{\mathrm{G}}(X)\cap\mathrm{G}(Y\neq0))\\
 & =\mathbb{P}(Y\neq0)H(X|Y\neq0)\\
 & =\mathbb{P}(Y\neq0)H(X).
\end{align*}
The blue cells add up to $I(Y;XY)$. Since $\mathcal{H}(\tilde{\mathrm{G}}(XY))$
is the sum of the blue cells and the red cell, we have
\[
H(XY)=\mathbb{P}(Y\neq0)H(X)+I(Y;XY),
\]
which completes the proof of the claim.
\end{example}

\medskip{}

To define the information-event diagram more formally, we can define
a generalization of the I-measure \cite{yeung1991new,yeung2012first},
called the \emph{IE-measure}, which is a signed measure $\mu_{\mathcal{H}}$
over the field of sets $\mathcal{S}_{n+m}$ generated by $n+m$ abstract
sets $\tilde{X}_{1},\ldots,\tilde{X}_{n},\tilde{E}_{1},\ldots,\tilde{E}_{m}$.
The field of sets $\mathcal{S}_{n+m}$ contains $2^{n+m}$ atoms in
the form of the formal intersection $(\bigcap_{i\in S}\tilde{X}_{i})\cap(\bigcap_{i\in[n]\backslash S}\tilde{X}_{i}^{c})\cap(\bigcap_{i\in S'}\tilde{E}_{i})\cap(\bigcap_{i\in[m]\backslash S'}\tilde{E}_{i}^{c})$
for $S\subseteq[n]$, $S'\subseteq[m]$, and contains all the $2^{2^{n+m}}$
measurable sets formed by unions of the atoms. The signed measure
$\mu_{\mathcal{H}}$ is given by
\begin{align*}
 & \mu_{\mathcal{H}}\bigg(\Big(\bigcap_{i\in S}\tilde{X}_{i}\Big)\cap\Big(\bigcap_{i\in[n]\backslash S}\tilde{X}_{i}^{c}\Big)\cap\Big(\bigcap_{i\in S'}\tilde{E}_{i}\Big)\cap\Big(\bigcap_{i\in[m]\backslash S'}\tilde{E}_{i}^{c}\Big)\bigg)\\
 & =\mathcal{H}\bigg(\Big(\bigcap_{i\in S}\tilde{\mathrm{G}}(X_{i})\Big)\cap\Big(\bigcap_{i\in[n]\backslash S}(\tilde{\mathrm{G}}(X_{i}))^{c}\Big)\cap\Big(\bigcap_{i\in S'}\mathrm{G}(E_{i})\Big)\cap\Big(\bigcap_{i\in[m]\backslash S'}(\mathrm{G}(E_{i}))^{c}\Big)\bigg),
\end{align*}
where $(\mathrm{G}(E_{i}))^{c}=\mathrm{G}(\Omega)\backslash\mathrm{G}(E_{i})$,
i.e., $\mu_{\mathcal{H}}$ is simply given by the composition $\mu_{\mathcal{H}}(S)=\mathcal{H}(\phi(S))$
for $S\in\mathcal{S}_{n+m}$, where $\phi:\mathcal{S}_{n+m}\to\mathrm{G}_{\mathrm{fin}}(\mathcal{F})$
is a homomorphism between the fields of sets $\mathcal{S}_{n+m}$
and $\mathrm{G}_{\mathrm{fin}}(\mathcal{F})$ (see Proposition \ref{prop:finadd_meas})
with $\phi(\tilde{X}_{i})=\tilde{\mathrm{G}}(X_{i})$ and $\phi(\tilde{E}_{i})=\mathrm{G}(E_{i})$.
The information-event diagram is then a diagram that shows the $2^{n+m}$
atoms of $\mathcal{S}_{n+m}$ and their measures. 

The IE-measure can potentially be used for automated proving of inequalities
involving entropy, mutual information and their conditional versions
given random variables and/or events, in a manner similar to \cite{yeung1996itip,yeung1997framework,ho2020proving,li2021automatedisit,yeung2021machine}.
This is left for future studies.

\medskip{}

\section{Harmonic Information Measure}

As an alternative to using Poisson processes in Definition \ref{def:poient_rv},
we can also define a similar measure using only an i.i.d. sequence.
\begin{defn}
\label{def:harent_rv}For $u\in\Omega$ and $\mathbf{v}=(v_{i})_{i\in\mathbb{N}}$,
$v_{i}\in\Omega$, and $X$ being a discrete random variable over
the probability space $(\Omega,\mathcal{F},\mathbb{P})$, the \emph{harmonic
information measure} of $X$ at $u$ with respect to $\mathbf{v}$
is
\[
\mathcal{R}_{u;\mathbf{v}}(X)=\sum_{i=1}^{\kappa_{u;\mathbf{v}}(X)-1}\frac{1}{i},
\]
where
\[
\kappa_{u;\mathbf{v}}(X)=\min\left\{ i\in\mathbb{N}:\,X(v_{i})=X(u)\right\} 
\]
is the first appearance time of $X(u)$ in the sequence $X(v_{1}),X(v_{2}),\ldots$.
\end{defn}
\medskip{}

Similar to Definition \ref{def:poient_gen}, we can also define the
harmonic information measure of a generalized information.

\medskip{}

\begin{defn}
\label{def:poient_gen-1}For a generalized information $A\in\mathrm{G}(\mathcal{F})$,
$u\in\Omega$ and $\mathbf{v}=(v_{i})_{i\in\mathbb{N}}$, $v_{i}\in\Omega$,
the \emph{harmonic information measure} of $A$ at $u$ with respect
to $\mathbf{v}$ is
\[
\mathcal{R}_{u;\mathbf{v}}(A)=\sum_{i=1}^{\infty}\frac{1}{i}\sum_{S\subseteq[i]}(-1)^{|S|+1}\mathbf{1}\big\{(u,v_{S_{(1)}},\ldots,v_{S_{(|S|)}})\in A\big\}
\]
where we assume $S=\{S_{(1)},\ldots,S_{(|S|)}\}$, $S_{(1)}<\cdots<S_{(|S|)}$.
The \emph{(average) harmonic information measures} are again defined
as
\[
\mathcal{R}_{u}(A)=\mathbb{E}\left[\mathcal{H}_{u;\mathbf{V}}(A)\right],
\]
where $\mathbf{V}=(V_{i})_{i\in\mathbb{N}}$ is an i.i.d. sequence
following $\mathbb{P}$, and
\[
\mathcal{R}(A)=\mathbb{E}\left[\mathcal{H}_{U;\mathbf{V}}(A)\right],
\]
where $U\sim\mathbb{P}$ is independent of $\mathbf{V}$.
\end{defn}
\medskip{}

It can be checked that $\mathcal{R}(\tilde{\mathrm{G}}(X))=H(X)$
(in nats), and most of the results on $\mathcal{H}$ in this paper
also applies to $\mathcal{R}$, with the exception of Propositions
\ref{prop:cond_event} and \ref{prop:cross_ent}. The problem about
$\mathcal{R}$ is that, unlike a $\mathbb{P}$-labeled Poisson process
where discarding points outside of the set $E$ would naturally yield
a $\mathbb{P}|_{E}$-labeled Poisson process, the procedure of discarding
entries in an i.i.d. sequence is much harder to describe mathematically
(one has to discard the entries and shift the indices of the remaining
entries accordingly).

\medskip{}

\section{Acknowledgement}

This work was supported in part by the Hong Kong Research Grant Council
Grant ECS No. CUHK 24205621. 

\medskip{}

\appendix

\subsection{Proof of Proposition \ref{prop:finadd_meas} \label{subsec:pf_finadd_meas}}

Given a random variable $W:\Omega\to\mathcal{W}$ where $\mathcal{W}$
is a finite set, and $B\subseteq2^{\mathcal{W}}$, define
\[
\tilde{\mathrm{G}}(W;B)=\left\{ (\omega_{1},\ldots,\omega_{k})\in\mathrm{G}(\Omega):\,\{W(\omega_{i}):1\le i\le k\}\in B\right\} .
\]
Note that $\tilde{\mathrm{G}}(W)=\tilde{\mathrm{G}}(W;\,2^{\mathcal{W}}\backslash\{\{w\}:w\in\mathcal{W}\})$,
and $\mathrm{G}(E)=\tilde{\mathrm{G}}(W;\{\{1\}\})$ where $W\in\{0,1\}$
is the indicator random variable of $E$. We first show that 
\[
\mathrm{G}_{\mathrm{fin}}(\mathcal{F})=\bigcup_{W,B}\{\tilde{\mathrm{G}}(W;B)\},
\]
where $W,B$ take values over measurable functions $W:\Omega\to\mathcal{W}$
where $\mathcal{W}$ is a finite set, and $B\subseteq2^{\mathcal{W}}$.
First we show that $\tilde{\mathrm{G}}(W;B)\in\mathrm{G}_{\mathrm{fin}}(\mathcal{F})$
for any $W,B$. Define $A_{|\mathcal{W}|},A_{|\mathcal{W}|-1},\ldots,A_{0}$
recursively as $A_{|\mathcal{W}|}=\emptyset$,
\[
A_{n-1}=\Big(A_{n}\cup\bigcup_{S\in B:\,|S|=n}\mathrm{G}(X^{-1}(S))\Big)\,\backslash\,\bigcup_{S\in2^{\mathcal{W}}\backslash B:\,|S|=n}\mathrm{G}(X^{-1}(S))
\]
for $n=|\mathcal{W}|,|\mathcal{W}|-1,\ldots,1$. For $(\omega_{1},\ldots,\omega_{k})\in\mathrm{G}(\Omega)$
with $S=\{W(\omega_{i}):1\le i\le k\}\in B$, we have $(\omega_{1},\ldots,\omega_{k})\in A_{|S|-1}$
by definition, and also $(\omega_{1},\ldots,\omega_{k})\in A_{n}$
for $n<|S|$ since $(\omega_{1},\ldots,\omega_{k})\notin\mathrm{G}(X^{-1}(S'))$
for $S'$ with $|S'|<|S|$. Similarly, for $(\omega_{1},\ldots,\omega_{k})\in\mathrm{G}(\Omega)$
with $S=\{W(\omega_{i}):1\le i\le k\}\notin B$, we have $(\omega_{1},\ldots,\omega_{k})\notin A_{n}$
for $n<|S|$. Hence $A_{0}=\tilde{\mathrm{G}}(W;B)$. Since $A_{0}$
is formed by finite union and difference between sets in the form
$\mathrm{G}(X^{-1}(S))$, we have $A_{0}=\tilde{\mathrm{G}}(W;B)\in\mathrm{G}_{\mathrm{fin}}(\mathcal{F})$.

Then we show the other direction that any entry of $\mathrm{G}_{\mathrm{fin}}(\mathcal{F})$
is in the form $\tilde{\mathrm{G}}(W;B)$. It suffices to show that
any complement and union of sets in the form $\tilde{\mathrm{G}}(W;B)$
is also in this form. We have
\[
\mathrm{G}(\Omega)\backslash\tilde{\mathrm{G}}(W;B)=\tilde{\mathrm{G}}(W;2^{\mathcal{W}}\backslash B),
\]
and
\begin{align*}
 & \tilde{\mathrm{G}}(W_{1};B_{1})\cup\tilde{\mathrm{G}}(W_{2};B_{2})\\
 & =\tilde{\mathrm{G}}\big((W_{1},W_{2});\,\{S\in2^{\mathcal{W}_{1}\times\mathcal{W}_{2}}:\,S_{1}\in B_{1}\;\mathrm{or}\;S_{2}\in B_{2}\}\big),
\end{align*}
where $S_{1}=\{x_{1}:\,(x_{1},x_{2})\in S\}$, $S_{2}=\{x_{2}:\,(x_{1},x_{2})\in S\}$
are projections of $S$. Hence any entry of $\mathrm{G}_{\mathrm{fin}}(\mathcal{F})$
is in the form $\tilde{\mathrm{G}}(W;B)$. 

Consider the case where $B=\{b\}$ (where $b\subseteq\mathcal{W}$)
is a singleton. We have
\begin{align*}
 & \mathcal{H}_{u;\mathbf{v}}(\tilde{\mathrm{G}}(W;\{b\}))\\
 & =\sum_{i=1}^{\infty}\Big(\log\frac{t_{i+1}}{t_{i}}\Big)\sum_{S\subseteq[i]}(-1)^{|S|+1}\mathbf{1}\big\{\{W(u)\}\cup\{W(v_{s}):\,s\in S\}=b\big\}.
\end{align*}
We study the sum $\sum_{S\subseteq[i]}\cdots$ above. Clearly, the
sum is zero if $W(u)\notin b$ or $b\nsubseteq\{W(u)\}\cup\{W(v_{s}):\,s\in[i]\}$,
so we assume $W(u)\in b$, $b\subseteq\{W(u)\}\cup\{W(v_{s}):\,s\in[i]\}$.
For notational simplicity, assume $\mathcal{W}=[n]$, $W(u)=1$ and
$b=[|b|]$. Let $K_{w}=\{j\in[i]:W(v_{j})=w\}$. Then $K_{2},\ldots,K_{|b|}\neq\emptyset$.
We have $\{W(u)\}\cup\{W(v_{s}):\,s\in S\}=b$ if $S\cap K_{w}\neq\emptyset$
for $2\le w\le|b|$, and $S\cap K_{w}=\emptyset$ for $w>|b|$. Hence,
\begin{align*}
 & \sum_{S\subseteq[i]}(-1)^{|S|}\mathbf{1}\big\{\{W(u)\}\cup\{W(v_{s}):\,s\in S\}=b\big\}\\
 & =\bigg(\sum_{S\subseteq K_{1}}(-1)^{|S|}\bigg)\bigg(\prod_{w=2}^{|b|}\sum_{S\subseteq K_{w},\,S\neq\emptyset}(-1)^{|S|}\bigg)\\
 & =\begin{cases}
(-1)^{|b|-1} & \mathrm{if}\;K_{1}=\emptyset\\
0 & \mathrm{if}\;K_{1}\neq\emptyset.
\end{cases}
\end{align*}
Therefore,
\begin{align*}
 & \mathcal{H}_{u;\mathbf{v}}(\tilde{\mathrm{G}}(W;\{b\}))\\
 & =\sum_{i=1}^{\infty}\Big(\log\frac{t_{i+1}}{t_{i}}\Big)\sum_{S\subseteq[i]}(-1)^{|S|+1}\mathbf{1}\big\{\{W(u)\}\cup\{W(v_{s}):\,s\in S\}=b\big\}\\
 & =(-1)^{|b|}\sum_{i=1}^{\infty}\Big(\log\frac{t_{i+1}}{t_{i}}\Big)\mathbf{1}\big\{\{W(u)\}\subseteq b\subseteq\{W(u)\}\cup\{W(v_{s}):\,s\in[i]\}\\
 & \quad\quad\quad\quad\quad\quad\quad\quad\quad\quad\quad\;\mathrm{AND}\;\{W(u)\}\cap\{W(v_{s}):\,s\in[i]\}=\emptyset\big\}\\
 & =(-1)^{|b|}\mathbf{1}\{W(u)\in b\}\log\frac{\min_{i:\,W(v_{i})=W(u)}t_{i}}{\min_{i:\,W(v_{i})=W(u)\;\mathrm{OR}\;b\subseteq\{W(u)\}\cup\{W(v_{s}):\,s\in[i]\}}t_{i}}.
\end{align*}
To bound $\mathcal{H}_{u;\mathbf{v}}(\tilde{\mathrm{G}}(W;\{b\}))$,
we have
\begin{align*}
 & (-1)^{|b|}\mathcal{H}_{u;\mathbf{v}}(\tilde{\mathrm{G}}(W;\{b\}))\\
 & \ge\mathbf{1}\{W(u)\in b\}\log\frac{\min_{i:\,W(v_{i})=W(u)}t_{i}}{\min_{i:\,W(v_{i})=W(u)}t_{i}}\\
 & \ge0,
\end{align*}
and
\begin{align*}
 & (-1)^{|b|}\mathcal{H}_{u;\mathbf{v}}(\tilde{\mathrm{G}}(W;\{b\}))\\
 & \le(-1)^{|b|}\mathbf{1}\{W(u)\in b\}\log\frac{\min_{i:\,W(v_{i})=W(u)}t_{i}}{\min_{i}t_{i}}.
\end{align*}
Taking expectation over a $\mathbb{P}$-labeled Poisson process $\mathbf{V}$,
we have
\begin{align*}
0 & \le(-1)^{|b|}\mathcal{H}_{u}(\tilde{\mathrm{G}}(W;\{b\}))\\
 & \;\;\le(-1)^{|b|+1}\mathbf{1}\{W(u)\in b\}\log\mathbb{P}(W^{-1}(u)),
\end{align*}
and
\begin{align*}
0 & \le(-1)^{|b|}\mathcal{H}(\tilde{\mathrm{G}}(W;\{b\}))\\
 & \;\;\le(-1)^{|b|+1}\mathbb{P}(W^{-1}(u))\log\mathbb{P}(W^{-1}(u)).
\end{align*}
Hence, $\mathcal{H}(\tilde{\mathrm{G}}(W;\{b\}))$ is well-defined
and finite. Since the limit of the sum of convergent sequences is
the sum of the limits, we know that $\mathcal{H}(\tilde{\mathrm{G}}(W;B))=\sum_{b\in B}\mathcal{H}(\tilde{\mathrm{G}}(W;\{b\}))$
is also well-defined and finite, and $\mathcal{H}$ is a finitely-additive
signed measure over $\mathrm{G}_{\mathrm{fin}}(\mathcal{F})$.

\medskip{}

\bibliographystyle{IEEEtran}
\bibliography{ref}

\end{document}